\documentclass[a4paper,fleqn,usenatbib,useAMS]{mnras}

\usepackage{graphicx}	
\usepackage{amsmath}	
\usepackage{amssymb}	
\usepackage{multicol}   
\usepackage{bm}		
\usepackage{pdflscape}	
\usepackage[normalem]{ulem}

\usepackage[T1]{fontenc}
\usepackage{ae,aecompl}

\usepackage{newtxtext,newtxmath}

\title[\textit{Planck's} view on the spectrum of the SZ effect]{\textit{Planck's} view on the spectrum of the Sunyaev--Zeldovich effect}

\author[Erler, Basu, Chluba \& Bertoldi]{Jens Erler$^{1}$\thanks{E-mail: \href{mailto:jens@astro.uni-bonn.de}{jens@astro.uni-bonn.de}}\thanks{Member of the International Max Planck Research School (IMPRS) for Astronomy and Astrophysics at the Universities of Bonn and Cologne.}, Kaustuv Basu$^{1}$, Jens Chluba$^{2}$, Frank Bertoldi$^{1}$
\\
$^{1}$Argelander-Institut f{\"u}r Astronomie, Universit{\"a}t Bonn, Auf dem H{\"u}gel 71, 53121 Bonn, Germany \\
$^{2}$Jodrell Bank Centre for Astrophysics, University of Manchester, Oxford Road, M13 9PL, U.K.}

\date{Last updated 2015 May 22; in original form 2013 September 5}

\pubyear{2018}

\begin{document}
\label{firstpage}
\pagerange{\pageref{firstpage}--\pageref{lastpage}}
\maketitle

\begin{abstract}
We present a detailed analysis of the stacked frequency spectrum of a large sample of galaxy clusters using \textit{Planck~} data, together with auxiliary data from the \textit{AKARI} and \textit{IRAS} missions. Our primary goal is to search for the imprint of relativistic corrections to the thermal Sunyaev--Zeldovich effect (tSZ) spectrum, which allow to measure the temperature of the intracluster medium. We remove Galactic and extragalactic foregrounds with a matched filtering technique, which is validated using simulations with realistic mock data sets. The extracted spectra show the tSZ signal at high significance and reveal an additional far-infrared (FIR) excess, which we attribute to thermal emission from the galaxy clusters themselves. This excess FIR emission from clusters is accounted for in our spectral model. We are able to measure the tSZ relativistic corrections at $2.2\sigma$ by constraining the mean temperature of our cluster sample to $4.4^{+2.1}_{-2.0} \, \mathrm{keV}$. We repeat 
the same analysis on a subsample containing only the 100 hottest clusters, for which we measure the mean temperature to be $6.0^{+3.8}_{-2.9} \, \mathrm{keV}$, corresponding to $2.0\sigma$. The temperature of the emitting dust grains in our FIR model is constrained to $\simeq 20 \, \mathrm{K}$, consistent with previous studies. 
Control for systematic biases is done by fitting mock clusters, from which we also show that using the non-relativistic spectrum for SZ signal extraction will lead to a bias in the integrated Compton parameter $Y$, which can be up to 14\% for the most massive clusters.
We conclude by providing an outlook for the upcoming CCAT-prime telescope, which will improve upon \textit{Planck~} with lower noise and better spatial resolution. 
\end{abstract}

\begin{keywords}
galaxies: clusters: general -- galaxies: clusters: intracluster medium -- cosmic background radiation -- cosmology: observations
\end{keywords}



\begingroup
\let\clearpage\relax
\endgroup
\newpage

\section{Introduction}

The Sunyaev--Zeldovich (SZ) effect is a spectral distortion of the cosmic microwave background (CMB) due to inverse Compton scattering of CMB photons by free electrons by the hot plasma found in clusters of galaxies. The effect was first described by \citet{Sunyaev70, Sunyaev72} and has been used extensively in the last two decades to detect and characterize galaxy clusters (e.g. \citealt{Hasselfield13, Planck_PSZE, Bleem15, Bender16}). 

The SZ signal is composed  of two distinct parts, the thermal SZ (tSZ) caused by the scattering of CMB photons by thermal electrons and the kinetic SZ (kSZ), which is due to scattering of CMB photons by a population of electrons that moves with a line-of-sight peculiar velocity $\varv_\mathrm{pec}$ in the rest frame of the CMB. Detailed reviews of the SZ effect are provided by \citet{Birkinshaw99} and \citet{Carlstrom02}. Given the dimensionless frequency \hbox{$x \equiv h\nu/(k_\mathrm{B}T_\mathrm{CMB})$}, the SZ signal can be expressed as an intensity shift relative to the CMB 
\begin{equation}
 \frac{\Delta I_\mathrm{SZ}}{I_\mathrm{0}} = h(x)\bigg[\underbrace{f(x, T_\mathrm{e}) \, y}_{\mathrm{tSZ}} - \underbrace{\tau_\mathrm{e} \, \left(\frac{\varv_\mathrm{pec}}{c}\right)}_\mathrm{kSZ}\bigg],
\end{equation}
where $y$ is the Compton $y$-parameter, a dimensionless measure of the line-of-sight integral of the electron pressure
\begin{equation}
y = \frac{\sigma_\mathrm{T}}{m_\mathrm{e}c^2}\int_\mathrm{l.o.s.} n_\mathrm{e} k_\mathrm{B} T_\mathrm{e} \, \mathrm{d}l.
\end{equation}
Here, $k_\mathrm{B}$ is the Boltzmann constant, $\sigma_\mathrm{T}$ is the Thomson cross-section, $m_\mathrm{e}$ is the electron rest mass, $c$ is the speed of light, $T_\mathrm{CMB}$ is the CMB temperature, \hbox{$I_0 = 2(k_\mathrm{B}T_\mathrm{CMB})^3/(hc)^2\approx 270 \, \mathrm{MJy \, sr^{-1}}$}, \hbox{$h(x) = x^4 \exp(x) / (\exp(x)-1)^2$} and \hbox{$\tau_\mathrm{e} = \sigma_\mathrm{T} \int n_\mathrm{e}(r) \, \mathrm{d}l$} is the optical depth of the plasma.
The function $f(x)$ describes the spectral shape of the tSZ effect
\begin{equation}
 f(x, T_\mathrm{e}) = \left( x \frac{\exp(x)+1}{\exp(x)-1} - 4 \right) + \delta_\mathrm{rel}(x, T_\mathrm{e}), 
\end{equation}
where $\delta_\mathrm{rel}(x, T_\mathrm{e})$ denotes relativistic corrections to the frequency spectrum of the tSZ \citep[e.g.,][]{Wright79, Rephaeli1995, Itoh98}, which arise from the high electron temperature of a few keV found in the intracluster medium (ICM) of galaxy clusters. These corrections (sometimes referred to as the relativistic SZ effect or the rSZ effect) 
can be efficiently computed using \hbox{{\small SZPACK}}\footnote{\url{www.Chluba.de/SZpack}} \citep{Chluba12, Chluba13}, which overcomes limitations of asymptotic expansions \citep{Challinor98, Itoh98, Sazonov98} and explicit tabulation schemes \citep[e.g.,][]{Nozawa00}. For the kSZ effect, we neglect relativistic corrections, which are well below the current sensitivity.

\begin{figure}
\includegraphics{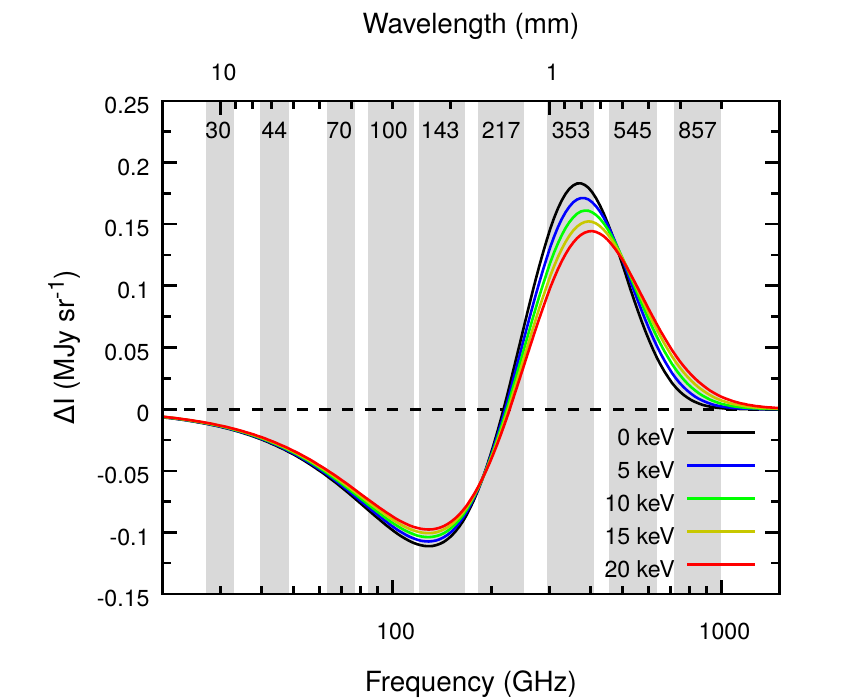}
\caption{Spectrum of the thermal SZ effect with relativistic corrections for a range of electron temperatures at fixed $y = 10^{-4}$. The grey bands indicate the nine \textit{Planck~} frequency bands with $\Delta \nu$/$\nu = 0.2$ for the three low-frequency instrument and $\Delta \nu$/$\nu = 0.3$ for the six high-frequency instrument channels.}
\label{fig:rsze_spectrum}
\end{figure}

In its non-relativistic approximation, the tSZ effect has a characteristic spectral shape independent of the plasma temperature, causing a decrement in intensity at frequencies below the tSZ 'null' at $\simeq 217.5 \, \mathrm{GHz}$ and an increment above. Taking into account relativistic corrections, the frequency spectrum becomes a function of the electron temperature. With increasing temperature, the tSZ 'null' shifts towards higher frequencies and the tSZ decrement and increment amplitudes decrease while the increment becomes wider (see \hbox{Fig.~\ref{fig:rsze_spectrum}}). For a massive galaxy cluster with $k_\mathrm{B}T_\mathrm{e} = 10 \, \mathrm{keV}$ the tSZ intensity at $353 \, \mathrm{GHz}$, for example, reduces by $13\%$. Accurate measurements of the spectral shape of the SZ spectrum would thus allow us to measure the $y$-weighted line-of-sight averaged ICM temperature of galaxy clusters (e.g. \citealt{Pointecouteau98}), allowing a more complete thermodynamic description without the need for additional density or temperature measurements from X-ray telescopes.

Since the SZ effect is a small distortion of the CMB, measuring weak changes in its spectrum at the level of a few per cent caused by relativistic effects or the similarly weak kSZ is very challenging and only recently have observations become sensitive enough. 
For example, \citet{Zemcov12} reported a 3$\sigma$ measurement of the shift of the SZ null using the \textit{Z}-spec instrument.
Under the assumption that the zero-shift is only caused by the relativistic  distortions (i.e. no kSZ), the authors constrained the temperature of the cluster \hbox{RX J 1347.5-1145} to $k_\mathrm{B}T_\mathrm{e} = (17.1 \pm 5.3) \, \mathrm{keV}$ . \citet{Prokhorov12} present a measurement of the line-of-sight temperature dispersion of the Bullet Cluster with observations of both the decrement and increment of the tSZ using data from ACBAR and \textit{Herschel}-SPIRE. Their analysis was later refined by \citet{Chluba13}, showing that no significant temperature dispersion could be deduced. In an attempt to measure the evolution of the CMB temperature, \citet{Hurier14} demonstrated that constraints on the electron temperature of a sample of clusters can be placed using data from the \textit{Planck~} satellite. 
More recently, \citet{Hurier16} claimed a high significance detection of the tSZ relativistic corrections based on a stacking analysis performed on large cluster samples using \textit{Planck~} data.

A major challenge for precision measurements of the electron temperature of galaxy clusters via the relativistic tSZ effect is far-infrared emission (FIR) that is spatially correlated with clusters and can affect measurements of the tSZ increment. Galaxy clusters are populated with galaxies, some of which form stars,  which then in turn heat up the dusty interstellar medium (ISM) of these galaxies, giving rise to thermal emission from warm dust grains. Although the star formation rates in most clusters are low, some are known to show exceptionally high star formation activity (e.g. \citealt{McDonald16}). 
This dusty galaxy contribution corresponds to the halo--halo clustering term of the cosmic infrared background (CIB) that is correlated with cluster positions (e.g., \citealt{Addison12}).
Individual CIB sources are also magnified by clusters through gravitational lensing, leading to spatially correlated increases in the CIB flux \citep{Blain98}. 
In addition to the unresolved galaxies, it has long been suspected that the ICM should contain large amounts of warm ($\simeq 20 \, \mathrm{K}$) dust grains, which are thought to be stripped from infalling galaxies by ram pressure and supernova winds (e.g. \citealt{Sarazin88}). The dust grains are then stochastically heated by collisions with hot
electrons from the ICM and re-emit the absorbed energy in the FIR \citep{Ostriker73, Dwek90}. In the ICM, dust grains can be destroyed by thermal sputtering \citep{Draine79}, but the grain lifetimes are highly uncertain and depend on the ICM density and temperature, as well as the size of the dust grains, but can reach several billion years in the outskirts of clusters \citep{Dwek92}. 
The actual amounts of dust grains and their lifetime in the ICM is speculative and only recently have dust grains been included in hydrodynamical simulations of galaxies \citep{McKinnon16, McKinnon17}.
All of the above contribute to an FIR excess observed at low resolution in stacked samples of clusters \citep{Montier05, Giard08, Planck_tSZE_CIB, Planck_cluster_dust}.
Besides these spatially correlated sources of FIR emission, the spatially uncorrelated contribution of diffuse Galactic foregrounds like synchrotron, free--free and thermal dust emission, as well as the stochastic CIB from extragalactic sources, has to be subtracted or modelled carefully in order to allow for precise measurements of the SZ spectral shape.

In this work, we present a detailed analysis of the SZ spectrum of a stacked sample of galaxy clusters as seen by the \textit{Planck~} satellite. We remove Galactic and extragalactic foregrounds with a spatial matched filtering approach and include an FIR component in our model of the observed cluster spectrum. We provide an estimate of the sample mean electron temperature as well as the average FIR emission from clusters. A major aspect of our work is a realistic simulation set-up with mock clusters with which we test our method and demonstrate a potential $Y$-bias in the \textit{Planck~} SZ measurements, resulting from the use of the non-relativistic tSZ spectrum. As an outlook, we compare \textit{Planck~} to the upcoming CCAT-prime\footnote{\url{http://www.ccatobservatory.org/}} telescope that will offer exciting observational possibilities like determining the SZ spectral shape for large number of clusters.

Our paper is structured as follows: Section~\ref{sec:Data} provides an overview over the maps and cluster catalogues used in this work. Section~\ref{sec:Method} describes our matched filtering and stacking methods that are tested on mock data in Section~\ref{sec:simulations}. Section~\ref{sec:Results} presents our results. In Section~\ref{sec:Discussion} we provide a discussion of our results as well as a comparison with some contemporary works and give an outlook to future experiments. Section~\ref{sec:Conclusion} provides a summary and concludes our analysis. 

Throughout this paper, we assume a flat $\Lambda$CDM cosmology with $\Omega_\Lambda = 0.7$, $\Omega_\mathrm{b} = 0.05$, $h = 0.7$ and $T_\mathrm{CMB} = 2.7255 \, \mathrm{K}$, while $E(z) \equiv H(z)/H_0 = (\Omega_\mathrm{m}(1+z)^3 + \Omega_\Lambda)^{1/2}$ is the redshift-dependent Hubble ratio. Unless noted otherwise, the quoted parameter uncertainties refer to the $68\%$ confidence interval. We made use of the {\small IDL} Astronomy Library \citep{Landsman93} and all-sky maps were processed with {\small HEALPIX} (v3.30; \citealt{Gorski05}).

\section{Data sets}
\label{sec:Data}

\subsection{\textit{Planck~} all-sky maps}

The main data used in our analysis are the all-sky microwave maps captured by the \textit{Planck~} satellite that were taken from the full data release in 2015 (R2.02; \citealt{Planck_overview}). \textit{Planck~} has observed the sky over a period of 4$\,$yr and delivered maps in nine different frequency bands with two main instruments. The low frequency instrument (LFI) observed the sky in three bands ranging from $30\, \mathrm{GHz}$ to $70\, \mathrm{GHz}$ and completed a total of eight all-sky surveys. \textit{Planck's} high frequency instrument (HFI; \citealt{Planck_HFI}) observed in six bands between $100\, \mathrm{GHz}$ and $857\, \mathrm{GHz}$ and completed five all-sky surveys before the depletion of the necessary coolant. With its wide frequency coverage, \textit{Planck~} allows to probe the entire spectrum of the SZ (see Fig.~\ref{fig:rsze_spectrum}), especially at the tSZ increment. For details on the time-ordered information (TOI) processing, the map-making process and calibration strategies we refer to the HFI 
and LFI papers. The main map characteristics are summarized in Table~\ref{tab:maps}. All maps are provided in the {\small HEALPIX} format with $N_\mathrm{side} = 2048$.
\begin{table}
\begin{center}
\tabcolsep=0.18cm
\begin{tabular}{cccc}
\hline
 $\nu \ (\mathrm{GHz})$ & $\lambda \ (\micron)$  & FWHM (arcmin) & Calibration uncertainty (\%) \\
\hline
\multicolumn{4}{{c}}{\textbf{\textit{Planck}}}  \\
70 & 4290 & 13.31 & 0.20 \\ 
100 & 3000 & 9.68 & 0.09 \\ 
143 & 2100 & 7.30 & 0.07 \\ 
217 & 1380 & 5.02 & 0.16 \\ 
353 & 850 & 4.94 & 0.78 \\ 
545 & 550 & 4.83 & 6.10 \\ 
857 & 350 & 4.64 & 6.40 \\ 
\multicolumn{4}{{c}}{\textbf{\textit{IRAS}/IRIS}}  \\
3000 & 100 & 4.30 & 13.5 \\ 
5000 & 60 & 4.00 & 10.4 \\ 
\multicolumn{4}{{c}}{\textbf{\textit{AKARI}}}  \\
3330 & 90 & 1.3 & 15.1 \\ 
\hline
\end{tabular}
\end{center}
\caption{Characteristics of the \textit{Planck~}, \textit{IRAS} and \textit{AKARI} all-sky maps used in this work. We adopt a covariance estimation approach similar to \citet{Soergel17} and assume that the calibration uncertainties between the \textit{Planck~} $\mathrm{70 \, \mathrm{GHz}}$ to $\mathrm{353 \, \mathrm{GHz}}$ and $\mathrm{545 \, \mathrm{GHz}}$ to $\mathrm{857 \, \mathrm{GHz}}$ channels are fully correlated.}
\label{tab:maps}
\end{table}
Our analysis uses all six HFI channels as well as the LFI $70 \, \mathrm{GHz}$ channel. The $30 \, \mathrm{GHz}$ and $44 \, \mathrm{GHz}$ LFI channels are not used due to their much lower angular resolution of 32 and 27$\,$arcmin, respectively, and their sensitivity to low-frequency synchrotron and free--free emission from both Milky Way and bight radio galaxies along the line of sight. We convert all maps up to $353 \, \mathrm{GHz}$ from units of $\mathrm{K_{CMB}}$ to $\mathrm{MJy \, sr^{-1}}$ with the unit conversion factors given in the \textit{Planck~} 2015 release explanatory supplement\footnote{We use the band-average unit conversion factors that can be found here: \url{https://wiki.cosmos.esa.int/planckpla2015/index.php/UC_CC_Tables}}. 

We adopt a covariance estimation approach similar to \citet{Soergel17}, who assume that the calibration uncertainties of channels that were jointly calibrated are fully correlated. The \textit{Planck~} LFI and HFI channels up to $353 \, \mathrm{GHz}$ where calibrated using the CMB dipole, while the two highest frequency maps were calibrated using Planets \citep{Planck_mapmaking}. In accordance with \citet{Soergel17}, we assume a conservative 1\% absolute calibration uncertainty for the channels up to $353 \, \mathrm{GHz}$ and 6\% for the two remaining channels. 

\subsection{\textit{IRAS} and \textit{AKARI} all-sky maps}

In addition to the \textit{Planck~} all-sky maps, we use auxiliary maps from the \textit{Infrared Astronomical Satellite} (\textit{IRAS}, \citealt{Neugebauer84}) and the \textit{AKARI} satellite \citep{Doi15} to constrain our spectral model at high frequencies. The main characteristics of the used maps are summarized in Table~\ref{tab:maps}.  

\textit{IRAS} performed the first all-sky survey in the mid-infrared and FIR in 1983 and delivered maps in four bands from $12 \, \micron$ to $100 \, \micron$. We make use of the reprocessed IRIS maps \citep{Miville05}, which offer improved calibration, zero level and de-striping, as well as better zodiacal light subtraction. Our analysis uses the IRIS $60 \, \micron$ and $100 \, \micron$ maps in the {\small HEALPIX} format with $N_\mathrm{side} = 2048$.  Both maps have similar resolution like the \textit{Planck~} high frequency bands but suffer from larger calibration uncertainties. 

The \textit{AKARI} satellite, also known as ASTRO-F, performed an all-sky FIR survey in four bands, covering wavelengths between $65 \, \micron$ and $160 \, \micron$. Compared to \textit{IRAS}, \textit{AKARI} offers a higher  angular resolution of $1$--$1.5\,$arcmin at a similar noise level. We only use the $90 \, \micron$ channel (WIDE-S) because it offers the lowest calibration uncertainties \citep{Takita15}. As for the other data sets, we obtained the \textit{AKARI} $90 \, \micron$ map in the {\small HEALPIX} format\footnote{The \textit{AKARI} maps can be downloaded in the {\small HEALPIX} format from the Centre d'Analyse de Donn\'{e}es Etendues (CADE, Paradis et al. \citeyear{Paradis12}): \ \url{http://cade.irap.omp.eu/dokuwiki/doku.php?id=start}} with $N_\mathrm{side} = 4096$ to account for the higher angular resolution.

\subsection{Galaxy cluster catalogues}

At the core of our analysis lies a stacking approach, which requires a large number of massive clusters for which the relativistic distortions of the tSZ spectrum can be significant. For this reason, the main cluster catalogue used in this study is the second \textit{Planck~} Catalogue of Sunyaev--Zeldovich sources (PSZ2; \citealt{Planck_PSZE2}), which provides the largest and deepest SZ-selected sample of galaxy clusters. The catalogue contains a total of 1653 detections, 1203 of which are confirmed galaxy clusters and 1094 have spectroscopic or photometric redshifts. The redshift range of the clusters is $0.01 \lesssim z \lesssim 0.97$ with a median redshift of $z_\mathrm{m} = 0.224$. We use the Union catalogue (R2.08), which combines the results of three distinct extraction algorithms. The MMF1 and MMF3 algorithms are based on matched multifiltering, a concept first proposed by \citet{Herranz02}, while the {\small POWELLSNAKES} (PwS) algorithm employs Bayesian inference. The provided estimates of the integrated 
Compton $y$-parameter within $5 \times R_{500}$ in the Union catalogue are taken from the algorithm that gave the highest signal-to-noise (S/N) detection for each individual cluster. Mass estimates are provided assuming the best-fit $Y$--$M$ scaling relation of \citet{Arnaud10} as a prior. The mass range of the galaxy clusters with known redshifts is  $7.8 \times 10^{13} \, \mathrm{M_\odot} \lesssim M_{500} \lesssim 1.6 \times 10^{15} \, \mathrm{M_\odot}$ with a median mass of $M_{500}^\mathrm{m} =  4.75 \times 10^{14} \, \mathrm{M_\odot}$.


\section{Method}
\label{sec:Method}

We search for the imprint of relativistic corrections to the tSZ by means of stacking multifrequency data for large samples of galaxy clusters. Since the relativistic corrections are expected to be weak ($\simeq 10\%$) even for massive and hot clusters, it is crucial to have high S/N data. Galactic foregrounds are reduced before the stacking of clusters by applying matched filters, tailored to the characteristic cuspy profile of galaxy clusters, to the all-sky maps. After filtering, the clusters are stacked within {\small HEALPIX} to avoid possible biases introduced by approximate projections.

\subsection{Matched filtering}
\label{sec:matched_filtering}

  Matched filtering is a technique that allows the construction of an optimal spatial filter to extract weak signals with a well-known spatial signature in the presence of much stronger foregrounds. Matched filtering was first proposed for the study of the kSZ by \citet{Haehnelt96} and was subsequently developed and generalized by \citet{Herranz02} and \citet{Melin06} for the extraction of the tSZ signal from multifrequency data sets like those delivered by the \textit{Planck~} mission. Matched filtering has since been adopted by the SPT, ACT and Planck Collaboration to extract the tSZ signal of clusters from their respective data sets (\citealt{Hasselfield13, Bleem15, Planck_PSZE2}).
  
\begin{figure}
\includegraphics{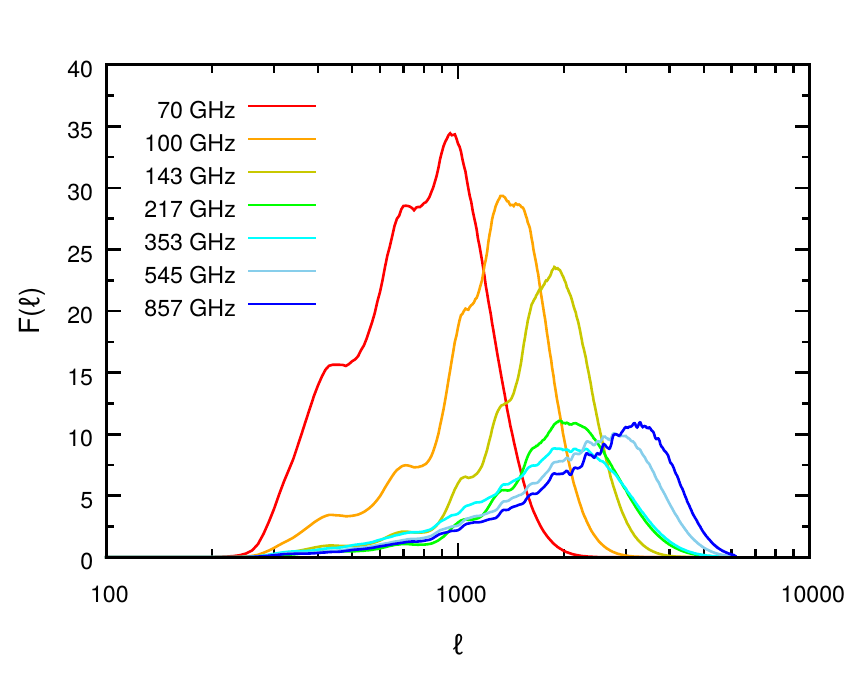}
\caption{Filter kernels for galaxy clusters in the \textit{Planck~} data. The filters were constructed following equation~(\ref{eq:filter}) using a core radius of $1 \ \mathrm{arcmin}$ and power spectra extracted from real data. For plotting purposes, the kernels were smoothed with a running average of $\Delta \ell = 50$. We use the unsmoothed filter kernels in our analysis.}
\label{fig:filters}
\end{figure}

  We apply our filter functions to the all-sky maps in spherical harmonic space to avoid using an approximate projection on to a flat-sky geometry. Assuming radial symmetry of the galaxy cluster profile (i.e. $m=0$) and following the approach presented in \citet{Schaefer06}, a matched filter $\Psi_{l0}$ can be constructed by minimizing the variance of the filtered field
  \begin{equation}
   \sigma^2 = \sum_\ell C_\ell \Psi_{\ell 0}^2,
  \end{equation}
  where $C_\ell$ is the power spectrum of the unfiltered map. At the same time, we demand the filtered field to be an unbiased estimator of the amplitude of the tSZ signal at the position of galaxy clusters. The latter condition can be rewritten as
  \begin{equation}
   \sum_\ell \tau_\mathrm{\ell 0} \Psi_{\ell 0} = 1,
  \end{equation}
 where $\tau_{\ell 0}$ are the $m=0$ spherical harmonic coefficients of the cluster profile. A solution to this optimization problem is given by 
  \begin{equation}
   \Psi_{\ell 0} = \left( \sum_\ell \frac{\tau_\ell^2}{C_\ell} \right)^{-1} \frac{\tau_\ell}{C_\ell}.
  \end{equation}
 Using the convolution theorem on the sphere, the spherical harmonic coefficients of the filtered map $a_\mathrm{\ell m}^\mathrm{filt}$ can be related to the ones of the unfiltered map $a_\mathrm{\ell m}^\mathrm{unfilt}$ by
  \begin{equation}
   a_\mathrm{\ell m}^\mathrm{filt} = \sqrt{\frac{4 \pi}{2\ell + 1}} \, \Psi_{\ell 0} \, a_\mathrm{\ell m}^\mathrm{unfilt} \equiv F_\ell \, a_\mathrm{\ell m}^\mathrm{unfilt}.
  \end{equation}
  We approximate the spatial profile of the cluster tSZ signal by a projected spherical $\beta$-model \citep{Cavaliere76}:
  \begin{equation}
   y(\theta) = y_0\left[1+\left(\frac{\theta}{\theta_c}\right)^2\right]^\frac{{1-3\beta}}{2},
  \end{equation}
  where $\theta_\mathrm{c}$ is the core radius. We set $y_0 = 1$ and adopt the commonly used value of $\beta =1$, for which an analytic spherical harmonic transform can be found \citep[e.g.,][]{Soergel17}
  \begin{equation}
   y_{\ell0} = 2 \pi \theta_c^2 K_0(\ell \theta_c),
  \end{equation}
  where $K_0$ is the modified Bessel function of the second kind. In order to account for the instrumental beam and the {\small HEALPIX} pixelization, we multiply with the beam and pixel window functions $B_{\ell}$ and $w_{\ell}$:
  \begin{equation}
   \tau_\ell = \sqrt{\frac{2\ell+1}{4 \pi}} \cdot {\tilde \tau}_{\ell 0} = \sqrt{\frac{2\ell+1}{4 \pi}} \cdot y_{\ell0} \cdot B_{\ell} \cdot w_{\ell}.
  \end{equation}
  All instrumental beams are assumed to be Gaussian with FWHMs as summarized in Table~\ref{tab:maps}. The final filters are therefore given by
  \begin{equation}
   F_\ell = \left[ \sum_\ell \frac{(2\ell+1){\tilde \tau}_\ell^2}{4 \pi C_\ell} \right]^{-1} \frac{{\tilde \tau}_\ell}{C_\ell}.
   \label{eq:filter}
  \end{equation}
  Fig.~\ref{fig:filters} shows the filter kernels for \textit{Planck~} and \textit{IRAS} data.
  A matched filter as the one defined here will provide an estimate of the deconvolved central $y$-parameter $y_0$.
  
  The $C_\ell$ are computed directly from the all-sky maps. To mitigate the strong foregrounds along the Galactic disc, the maps are multiplied with a smoothed ($2^\circ$) 40\% Galactic mask. To prevent contamination of the results by large-scale residuals, an exponential taper is applied to the filters at scales $\ell < 300$. In order to stack the extracted tSZ signal amplitudes of different clusters, we bin them according to their apparent size
  and match the core radius used to compute the filter functions to each subsample. We find that good results can be obtained with 11 size-bins between $\theta_c^\mathrm{min} = 0.25'$ and $\theta_c^\mathrm{max} = 3'$ with $\theta_\mathrm{c} = 0.2 \, \theta_{500}^\mathrm{m}$, where $\theta_{500}^\mathrm{m}$ is the median $\theta_{500}$ for each subsample.
  
\subsection{Sample selection} 

\begin{figure}
\includegraphics[width=\columnwidth]{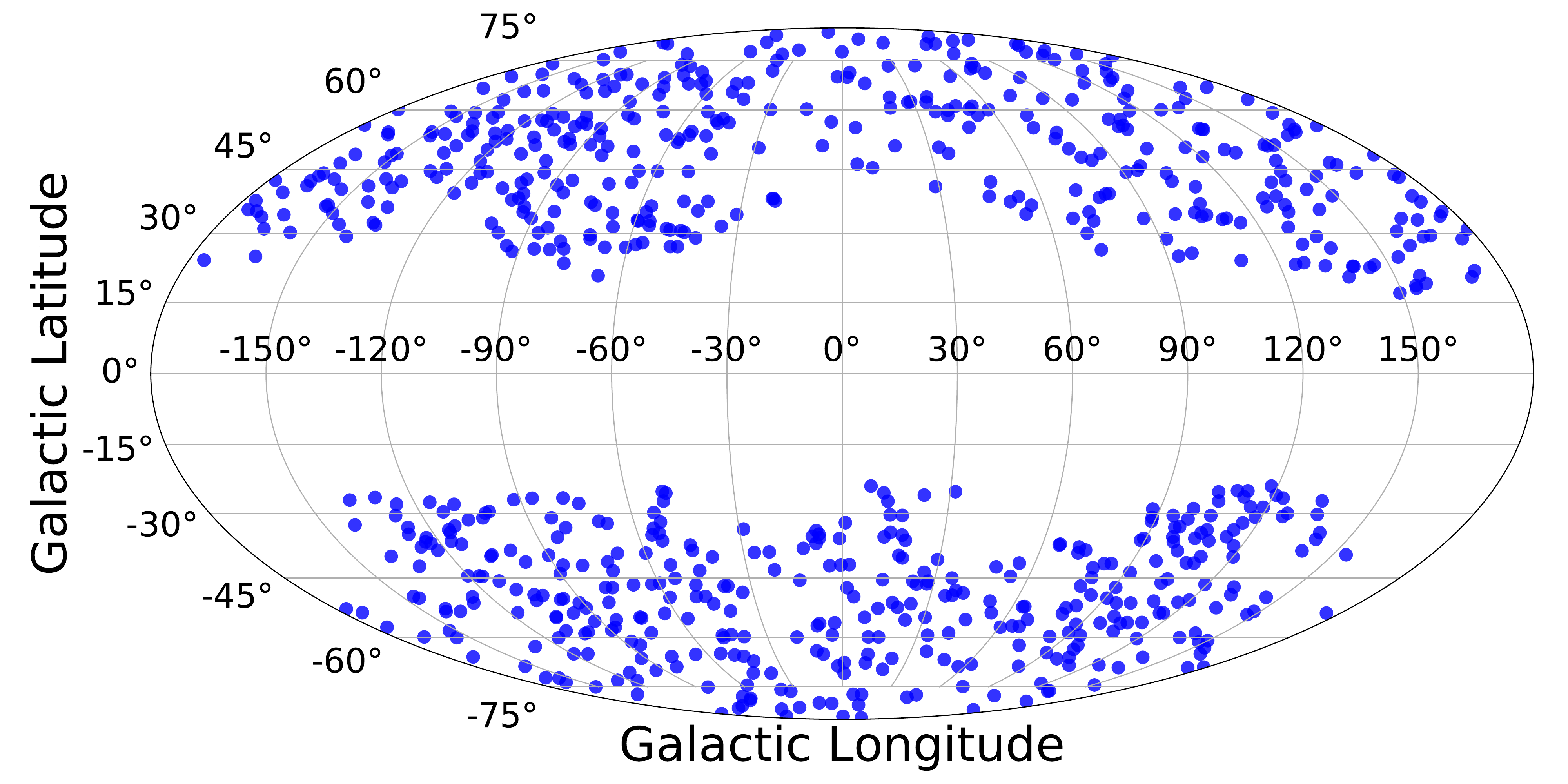}
\caption{Mollweide projection of the sky in Galactic coordinates showing the subsample of the second \textit{Planck~} cluster catalogue used in this work. The cluster-free central part of the image traces the Galactic mask used for cluster selection. We also flag all clusters with a known low-frequency point sources within a radius of $10 \, \mathrm{arcmin}$ from the cluster centre and exclude large low-redshift systems (see the main text).}
\label{fig:sample}
\end{figure}

\begin{figure*}
\includegraphics{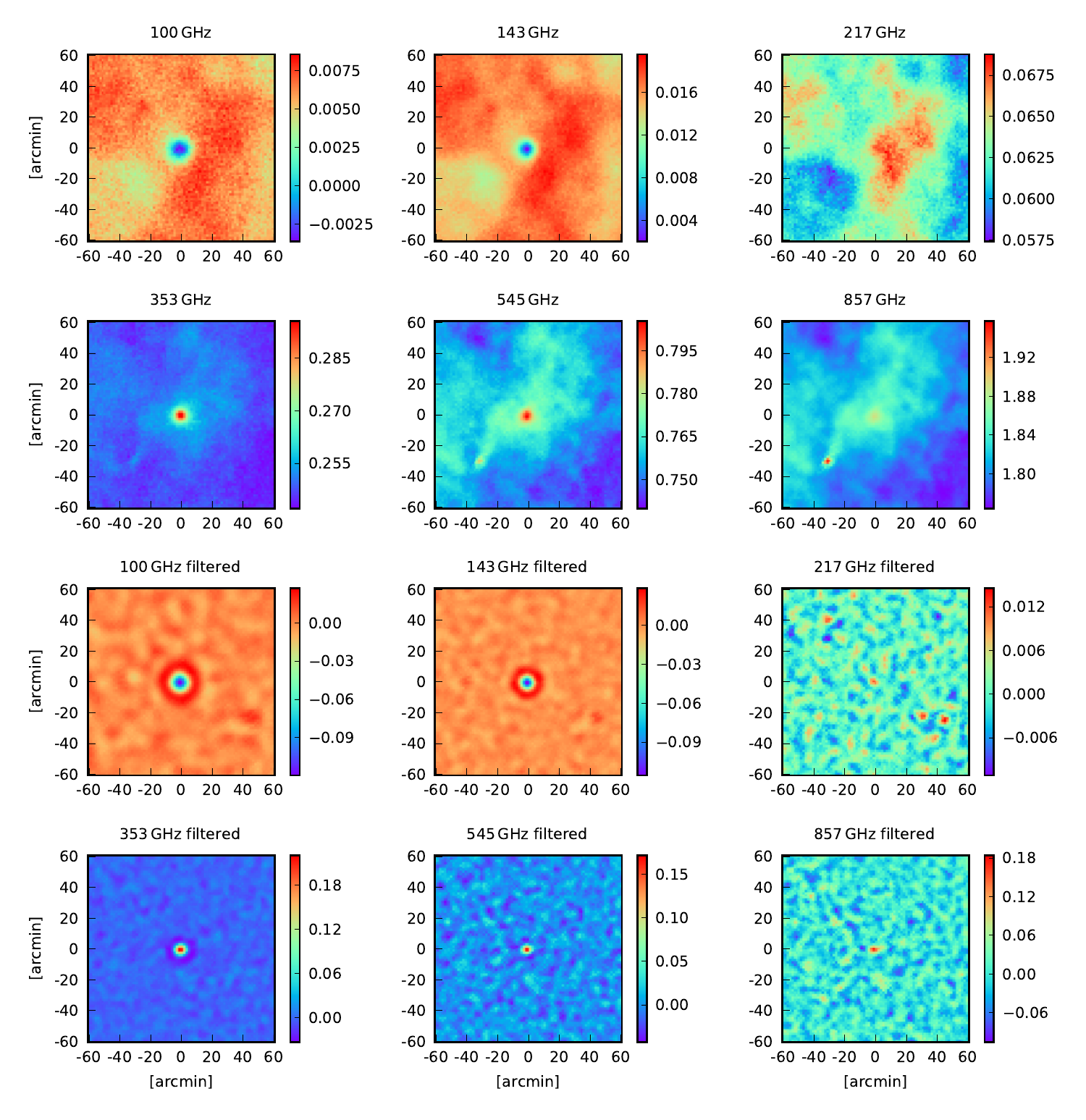}
\caption{Stacked \textit{Planck~} HFI maps for our final sample of 772 clusters. All fields are $2^\circ \times 2^\circ$ and the units are $\mathrm{MJy \, sr^{-1}}$. The panels in the upper two rows were created by smoothing all HFI channels to a common resolution of $9.66 \, \mathrm{arcmin}$ and then stacking the cluster positions without further foreground or background removal. Although the stacking procedure averaged out most contaminants, there are still large inhomogeneities present in the maps. In contrast, the panels shown in the two lower rows were created by stacking the HFI data after matched filtering, which removed most contaminants with great efficiency. We note that we stack the maps directly in {\small HEALPIX} for our spectral analysis and show these panels only for the purpose of illustration.}
\label{fig:snapshots}
\end{figure*}

In order to avoid the strong Galactic foregrounds along the Galactic plane, we exclude galaxy clusters that fall within a 40\% Galactic dust mask. Some galaxy clusters are also known to host bright radio galaxies that can bias measurements of the tSZ decrement (e.g. \citealt{Lin07}). To avoid the brightest sources, we remove all clusters from our sample that have a known point source detected within a radius of $10 \, \mathrm{arcmin}$ from the cluster centre. For this purpose, we use the \textit{Planck~} catalogue of compact sources in its second iteration ($30 \, \mathrm{GHz}$ to $217 \, \mathrm{GHz}$, \citealt{Planck_point_catalog}) and include weak detections. These two steps reduce the size of our sample to 821 clusters. We furthermore exclude clusters with $\theta_{500} > 15 \, \mathrm{arcmin}$ in order to keep the number of size bins used in the matched filtering step low. By doing so, we exclude an additional 49 low-redshift clusters and are left with our final sample of 772 clusters, the positions of 
which are shown in Fig.~\ref{fig:sample}. Our cluster sample has a median redshift of 0.23, the mean redshift is 0.27 and the mean cluster mass is $\langle M_{500} \rangle = 4.8 \times 10^{14} \, \mathrm{M_\odot}$ with a standard deviation of $\sigma_{M_{500}} = 1.9\times 10^{14} \, \mathrm{M_\odot}$.
The stacked cluster sample is shown in Fig.~\ref{fig:snapshots} both without and with foreground-removal at the \textit{Planck~} HFI frequencies, highlighting the effectiveness of our matched filtering technique.

We use the $M$--$T$ scaling relation given by \citet{Reichert11} to obtain an estimate of the X-ray spectroscopic electron temperature of the clusters in our sample: 
\begin{equation}
    \frac{M_{500}}{10^{14} \, \mathrm{M_\odot}} = (0.291 \pm 0.031) \, \left(\frac{k_\mathrm{B}T_\mathrm{X}}{\mathrm{keV}} \right)^{1.62 \pm 0.08} E(z)^{-1.04 \pm 0.07}.
    \label{eq:M-T_relation}
\end{equation}
The error on the estimate of sample-average temperature is obtained via a Monte Carlo technique taking into account both the scaling relation uncertainties and the quoted mass errors in the \textit{Planck~} catalogue. This estimate of spectroscopic temperature is used to compare against the $\langle T_\mathrm{SZ}\rangle$ values as obtained from our tSZ spectral analysis. For example, we find a sample-average (mass-weighted) X-ray  temperature $k_\mathrm{B} \langle T_\mathrm{X} \rangle = (6.91\pm 0.08) \, \mathrm{keV}$ and sample standard deviation $k_\mathrm{B} \sigma_{T_\mathrm{X}} = 2.13 \, \mathrm{keV}$ for the full sample of 772 clusters.

In addition to our full sample, we select a subsample containing the 100 hottest clusters by employing the same $M$--$T$ scaling relation. This subsample thus contains the most massive clusters from our original sample, with a mean mass of \hbox{$\langle M_{500} \rangle = 7.9 \times 10^{14} \, \mathrm{M_\odot}$}, and a higher mean redshift of $0.46$ and a median of $0.45$. The sample-average mass-weighted spectroscopic temperature  is $k_\mathrm{B} \langle T_\mathrm{X} \rangle = (8.54\pm 0.16) \, \mathrm{keV}$ with a sample dispersion of \hbox{$k_\mathrm{B} \sigma_{T_\mathrm{X}} = 1.57 \, \mathrm{keV}$}. This sample allows us to test for a stronger relativistic tSZ signal with the drawback of a reduced sample size.

\subsection{Data modelling}

After matched filtering, the extracted spectra will be free of spatially uncorrelated Galactic and extragalactic foregrounds, thus we only model the expected signal from the galaxy clusters. We fit a two-component model to the data that is the sum of a tSZ spectrum with relativistic corrections and a model for the expected FIR emission from galaxy clusters.

We compute the tSZ spectrum using the {\small SZPACK} code in its {\tt `COMBO'} runmode, which delivers accurate results up to very high electron temperatures of $75 \, \mathrm{keV}$ by combining asymptotic expansions and improved pre-computed basis functions \citep{Chluba12}\footnote{Other fitting formulae (e.g. \citealt{Nozawa00}) show noticeable artefacts in the spectrum that are avoided with {\small SZPACK}.}. The instrumental bandpass is accounted for by adopting the approach presented by the \citet{Planck_HFI}
\begin{equation}
    \Delta\tilde{I}_\mathrm{tSZ}(x, T_\mathrm{e}) = y \, I_0 \, \frac{ \int \mathrm{d}\nu \, \tau(\nu) \, h(x) \, f_\mathrm{rel}(x, T_\mathrm{e})}{\int \mathrm{d}\nu \, \tau(\nu) \left( \frac{\nu_\mathrm{c}}{\nu} \right)}, 
\label{eq:bandpass}
\end{equation}
where $\nu_\mathrm{c}$ denotes the band central frequency and $\tau(\nu)$ the bandpass transmission\footnote{The bandpass transmission tables can be found here: \url{http://irsa.ipac.caltech.edu/data/Planck/release_1/ancillary-data/}} at the frequency $\nu$. Table~\ref{tab:SZspec} provides the bandpass-corrected tSZ spectrum with relativistic corrections for a range of temperatures. 
At the given range of cluster temperatures in our sample, fitting the extracted spectrum of the stacked clusters with a tSZ spectrum will provide an estimate of the sample-average central $y$-parameter $\langle y_0 \rangle$ and the pressure-weighted average electron temperature (e.g. \citealt{Hansen04})
\begin{equation}
 T_\mathrm{SZ} \approx \langle T_\mathrm{e} \rangle_{P_\mathrm{e}} = \frac{\int n_\mathrm{e} T_\mathrm{e}^2 \, \mathrm{d}l}{\int n_\mathrm{e}T_\mathrm{e} \, \mathrm{d}l}.
\end{equation}
We choose to model the FIR emission from galaxy clusters with a modified blackbody 
\begin{equation}
 \tilde{I}_\mathrm{FIR}(\nu) = \mathrm{CC}(\beta_\mathrm{Dust}, T_\mathrm{Dust}) \, A_\mathrm{Dust} \, \left(\frac{\nu}{\nu_0}\right)^{\beta_\mathrm{Dust}} \, B_\nu(T_\mathrm{Dust}),
 \label{eq:mbb}
\end{equation}
where $A_\mathrm{Dust}$, $T_\mathrm{Dust}$, and $\beta_\mathrm{Dust}$ are the dust amplitude, temperature, and spectral index, respectively, $\nu_0 = 857 \, \mathrm{GHz}$, \hbox{$B_\nu = 2\pi h \nu^3 / c^2 \, (\exp(h\nu/k_\mathrm{B}T_\mathrm{Dust})-1)^{-1}$} is Planck's law and $\mathrm{CC}$ denotes frequency-specific colour corrections
\begin{equation}
   \mathrm{CC}(\beta_\mathrm{Dust}, T_\mathrm{Dust}) = \frac{ \int \mathrm{d}\nu \, \tau(\nu) \left( \frac{\nu^{\beta_\mathrm{Dust}} \, B_\nu(\nu, T)}{\nu_{c}^{\beta_\mathrm{Dust}} \, B_\nu(\nu_c, T)} \right)}{\int \mathrm{d}\nu \, \tau(\nu) \left( \frac{\nu_\mathrm{c}}{\nu} \right)}.
\end{equation}
For convenience, we recast equation~(\ref{eq:mbb}) to
\begin{equation}
\tilde{I}_\mathrm{FIR}(\nu) = \mathrm{CC} \, A_\mathrm{Dust}^{857} \, \left(\frac{\nu}{\nu_0}\right)^{\beta_\mathrm{Dust}+3} \frac{\exp(h\nu_0/(k_\mathrm{B}T_\mathrm{Dust}))-1}{\exp(h\nu/(k_\mathrm{B}T_\mathrm{Dust}))-1},
\end{equation}
and report the measured FIR intensity at $857 \, \mathrm{GHz}$ as the amplitude $A_\mathrm{Dust}^{857}$.
We account for the redshift distribution of our cluster sample by computing the FIR model at each specific cluster redshift and averaging the obtained values. The obtained parameter values are thus given in the rest frame of the source.

\begin{figure}
\includegraphics[]{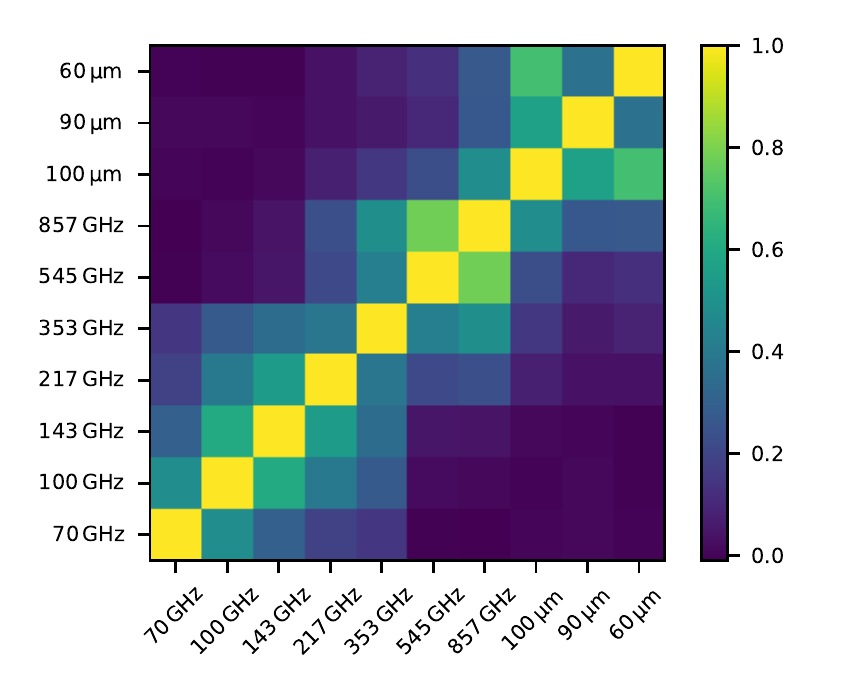}
\caption{Channel-to-Channel correlation matrix used for the data modelling. It is composed of a statistical and a systematic component. The statistical component is constrained by stacking 772 random positions across the sky outside of the same Galactic mask used for sample selection and repeating this exercise 10,000 times. The systematic component contains the calibration uncertainties of the instruments.}
\label{fig:correlation_matrix}
\end{figure}

Finally, we fit our data in a Bayesian approach by constraining the posterior probability distribution of our model parameters $\bm{p}$ using Markov Chain Monte Carlo (MCMC) sampling with
\begin{equation}
 P(\bm{p}|\bm{I}_\nu) \propto  P(\bm{I}_\nu|\bm{p}) \, P(\bm{p}),
\end{equation}
where $\bm{I}_\nu$ is the measured sample-average of specific intensities after matched filtering, $P(\bm{I}_\nu|\bm{p})$ is the likelihood function and $P(\bm{p})$ is the prior. We restrict the electron temperature to values \hbox{$0 \, \mathrm{keV} < T_\mathrm{e} < 75 \, \mathrm{keV}$} in accordance to {\small SZPACK}'s {\tt `COMBO'} runmode. Note that the sample-average temperature of the clusters should lie well within this range. We assume a flat positive prior on the remaining model parameters and a Gaussian likelihood that can be written as
\begin{equation}
\ln P(\bm{I}_\nu|\bm{p}) = -0.5 \, [\bm{I}_\nu(\bm{p})-\langle \bm{I}_\nu \rangle]^\mathrm{T} \mathbfss{C}^{-1} [\bm{I}_\nu(\bm{p})-\langle \bm{I}_\nu \rangle].
\end{equation}
The frequency-to-frequency covariance matrix $\mathbfss{C}$ is estimated by stacking 772 uniformly distributed random positions across the sky, excluding the area that falls into the Galactic mask used for sample selection. This step is repeated $10^4$ times, providing a large number of noise realizations for the covariance estimation. In this process, we account for the size binning of the clusters. This statistical component of the covariance matrix is then combined with the systematic part resulting from the instrumental calibration uncertainties. The corresponding correlation matrix is shown in \hbox{Fig.~\ref{fig:correlation_matrix}}.

We draw samples from the posterior probability distribution using an implementation of the Metropolis Hastings algorithm \citep{Metropolis53, Hastings70} and report the marginalized two-dimensional (2D) and one-dimensional (1D) posterior distributions. We ensure convergence by comparing the results of multiple chains that start from random positions in the parameter space.

\section{Simulations}
\label{sec:simulations}
\subsection{Simulation set-up}
\label{sec:simulation_set-up}

In order to test our filtering pipeline and data modelling procedure before applying it to real data, we validate it using realistic all-sky mock data. We use the CMB and Galactic synchrotron, free--free and thermal dust maps provided by the \citet{Planck_foregrounds} that were extracted with the Bayesian {\small COMMANDER} analysis framework. The {\small COMMANDER} synchrotron and free--free maps are provided at a low {\small HEALPIX} resolution of $N_\mathrm{side} = 256$ and are upgraded to $N_\mathrm{side} = 2048$ in spherical harmonic space to avoid pixelization artefacts. Thermal dust maps are provided at both low- and high- {\small HEALPIX} resolution. We use the $N_\mathrm{side} = 2048$ dust amplitude and $\beta_\mathrm{Dust}$ maps and upgrade the low-resolution dust temperature map to $N_\mathrm{side} = 2048$. Note that the same upgraded map was used as a prior during the creation of the $N_\mathrm{side} = 2048$ dust maps. The maps are scaled to \textit{Planck~} and \textit{IRAS} frequencies from $70 \, \mathrm{
GHz}$ to $5 \, \mathrm{THz}$ using the SEDs employed by the \citet{Planck_foregrounds}. We do not simulate \textit{AKARI} data due to a lack of high-resolution templates. However, testing our pipeline on \textit{Planck~} and \textit{IRAS} mock data is sufficient for our purposes.

The tSZ signal from clusters of galaxies is simulated by line-of-sight projection of a generalized Navarro-Frenk-White (GNFW) pressure profile \citep{Nagai07} with the mass-dependent parametrization given by \citet{Arnaud10}:
\begin{equation}
 \frac{P_\mathrm{e}(r)}{\mathrm{keV \, cm^{-3}}} = 1.65 \times 10^{-3} E(z)^{8/3} \left(\frac{M_{500}}{3\times10^{14} \, \mathrm{\mathrm{M_\odot}}}\right)^{0.79} p\left( \frac{r}{r_{500}} \right),
\end{equation}
where $p(r/r_{500})$ is the so-called "universal" shape of the cluster pressure profile:
\begin{equation}
 p(r) = \frac{P_0}{\left(c_{500} \frac{r}{R_{500}} \right)^\gamma \left[1+\left(c_{500} \frac{r}{R_{500}} \right)^\alpha \right]^{(\beta-\gamma)/\alpha}},
\end{equation}
with \hbox{$(P_0,c_{500},\gamma,\alpha,\beta) = (8.403,1.177,0.3081,1.0510,5.4905)$} as the best-fit values reported by \citet{Arnaud10}. We project the model along a series of concentric isothermal shells with $0.01 \, R_{500} < r < 3.5 \, R_{500}$ and $\Delta r = 0.1 \, R_{500}$, and assume the electron temperature to follow the profile given by \citet{Vikhlinin06}:
\begin{equation}
  \frac{T(r)}{T_\mathrm{mg}} = 1.35 \frac{\left(\frac{r}{0.045R_{500}} \right)^{1.9} + 0.45}{\left(\frac{r}{0.045R_{500}}\right)^{1.9} + 1} \frac{1}{\left[1+\left(\frac{r}{0.6R_{500}}\right)^2\right]^{0.45}},
\end{equation}
where $T_\mathrm{mg} = 0.9 T_\mathrm{X}$ accounts for the lower temperature due to weighting with the gas mass.
The tSZ signal is computed for each shell according to its temperature and $y$ parameter and the total signal for each cluster is given by the stack of all shells. Likewise, we also compute the optical depth $\tau_\mathrm{e}$ of each cluster shell by shell, as
\begin{equation}
 \tau_{\mathrm{e},i} = \frac{y_i  m_\mathrm{e} c^2}{k_\mathrm{B}T_{\mathrm{e},i}}
\end{equation}
and then estimate the total $\tau_\mathrm{e}$ by stacking all the shells. The $\tau_\mathrm{e}$ values are used to estimate the residual kSZ signal after stacking (Section \ref{sec:kSZ}).

To ensure similar signal strength to the real data, we adopt the cluster masses and redshifts from the previously described cluster sample but assign new random sky coordinates outside of a Galactic mask to each cluster to avoid placing them on top of spatially correlated artefacts in the foreground maps. These artefacts result from a lack of an SZ model during the foreground modelling and can introduce a bias in our parameter constraints. Randomizing the sky coordinates of the clusters allows us to obtain multiple foreground realizations with only one set of foreground templates.

We simulate the FIR emission of galaxy clusters by assuming a constant dust-to-gas mass ratio $M_\mathrm{Dust}$/$M_\mathrm{Gas} = 10^{-4}$ for all clusters as well as a modified blackbody spectral energy distribution (SED) with $T_\mathrm{Dust} = 20 \, \mathrm{K}$ and $\beta_\mathrm{Dust} = 1.5$, which are typical values found for the ISM of nearby galaxies and are consistent with the values reported by the \citet{Planck_cluster_dust}.
The amplitude of the SED can be related to the dust mass following the approach of \citet{Hildebrand83}
\begin{equation}
 A_\mathrm{Dust} = \frac{\kappa_\nu M_\mathrm{Dust}}{D_\mathrm{L}^2 \Omega}.
\end{equation}
where $\kappa_\nu$ is the mass absorption coefficient, $D_\mathrm{L}$ is the angular diameter distance, and $ \Omega = \pi (3\theta_{500})^2 / D_\mathrm{A}^2$ is the solid angle of the emitting region. We adopt the mass absorption coefficient reported by \citet{Draine03}, $\kappa_{850\micron} = 0.0383 \, \mathrm{m^2 \, kg^{-1}}$, which was also used by the \citet{Planck_cluster_dust}. Furthermore, the spatial profile of the FIR emission is assumed to follow a $\beta$-model with $\beta = 1$ and $r_\mathrm{c} = 0.2 \, R_{500}$. The \citet{Planck_tSZE_CIB} found that the FIR emission follows a broader profile compared to the tSZ signal, but its exact radial profile remains unknown. Since we reject clusters with known low-frequency point sources, i.e. radio galaxies, during the sample selection process, we do not include a radio-source component in our cluster simulations.

The foreground and cluster maps are then convolved with the instrument beams, which we approximate as circular Gaussians with FWHM as listed in Table~\ref{tab:maps}. 

We add an estimate of the instrumental noise to each map, which is obtained by computing the half difference \hbox{$(I_\nu^\mathrm{ring 1} - I_\nu^\mathrm{ring 2})/2$} of the two half-ring maps for every \textit{Planck~} channel, each of which only contains half of the stable pointing period data. Since no equivalent \textit{IRAS}/IRIS maps are available, we use white noise maps for which we adopt the global noise level found in the IRIS maps of $\sigma_\mathrm{60 \, \micron} = 0.03 \, \mathrm{MJy \, sr^{-1}}$ and $\sigma_\mathrm{100 \, \micron} = 0.06 \, \mathrm{MJy \, sr^{-1}}$.

{We have neglected the contribution from extragalactic dusty point sources in preparing our simulation set-up. These point sources constitute the CIB, with both one-halo (Poisson) and two-halo terms contributing to the relevant scales of tens of arcminutes (e.g., \citealt{Bethermin17}). The one-halo or Poisson term of the CIB acts as an additional source of thermal noise affecting the high-frequency bands, but otherwise is uncorrelated with the cluster location and will be filtered away by the matched-filtering technique. Hence in our simulations we have an under-estimation of the noise at the \textit{IRAS} frequencies as well as the highest frequency \textit{Planck~} channels, but this is not expected to result in any biases in the recovered model parameters. The contribution of the two-halo term in the CIB will be similar to the FIR component from the clusters already included in our analysis, barring a few extremely bright objects that will be flagged in a process similar to the 
cleaning of our real cluster sample. 

Other foreground components excluded from our simulations are mainly the Galactic CO and anomalous microwave emission (AME), as well as the Galactic and extragalactic radio point sources. Their aggregate contribution is expected to be small given our use of HFI-only data plus sky masking and sample cleaning methods (the all-sky mock data are filtered with the same pipeline as the real data). 
Adding these subdominant foregrounds can only be expected to make the parameter uncertainties marginally worse, hence their exclusion is not a concern while testing the robustness of our filtering pipeline. 

\subsection{Method validation with mock data}

Before applying our matched filtering pipeline to real data, it is important to assess if it allows an unbiased estimation of the cluster properties. In order to test our method, we simulate a total of 30 mock data sets and pass them through the same filtering and analysis pipeline as the real data. The obtained constraints on the tSZ parameters for all 30 data sets are shown in Fig.~\ref{fig:bias_sz}, while Fig.~\ref{fig:results_simulations} shows the extracted spectrum and model fitting as well as parameter constrains for one exemplary mock data set. 

\begin{figure}
\includegraphics{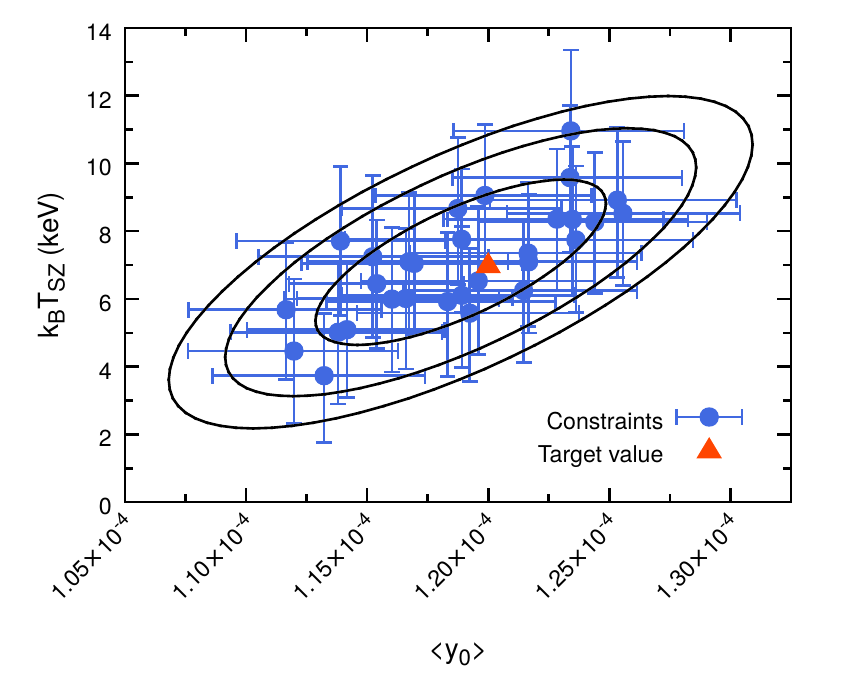}
\caption{Constraints on the tSZ parameters obtained from simulations with 30 different foreground realizations, achieved by randomizing the cluster coordinates. The true sample-average $y$ and $T_\mathrm{SZ}$ is indicated by the red triangle. The black contours indicate the 68, 95, and 99\% confidence intervals. This result suggests that our matched filtering and spectral fitting pipeline allows an unbiased measurement of the tSZ parameters.}
\label{fig:bias_sz}
\end{figure}

We find that the individual simulation constraints tightly scatter around the expected values that were derived directly from the simulated cluster data. This result demonstrates that an unbiased measurement of the sample-average $y_0$ and $T_\mathrm{SZ}$ can be achieved in a matched filtering approach with size binning. Similar results are found for the three parameters of the cluster FIR model. Assuming a different pressure profile like the best-fit GNFW model presented by the \citet{Planck_pressure} for our mock clusters while keeping our filter profile unchanged results in a bias in $y_0$ but not in $T_\mathrm{SZ}$. This bias can be avoided by choosing a different core radius $\theta_\mathrm{c} = 0.23 \, \theta_{500}^\mathrm{m}$ to construct our filters. The temperature $T_\mathrm{SZ}$ is insensitive to small differences between the true cluster shape and the assumed model for filtering since the mismatch will be the same across all frequencies. Differences in spatial resolution across the 
frequency bands could introduce a bias to the SZ spectral shape, but our tests suggest such distortions to be insignificant.

We also filter the mock data with a lower number of size bins that leads to an under-estimation of $y_0$ and overestimation of $T_\mathrm{SZ}$. Note that we do not test for potential biases due to cluster asphericity, which is a well-known problem in modelling individual objects \citep{Piffaretti03} but is not expected to cause a significant biases when stacking a large number of sources.

\begin{figure*}
\centering

\includegraphics[width = 0.55\textwidth]{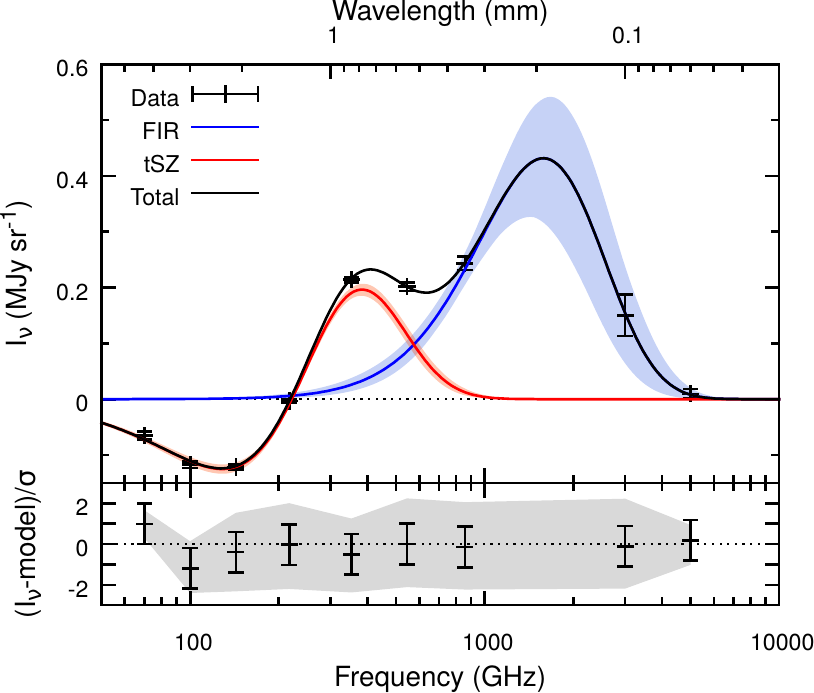}
\includegraphics[width = 0.7\textwidth]{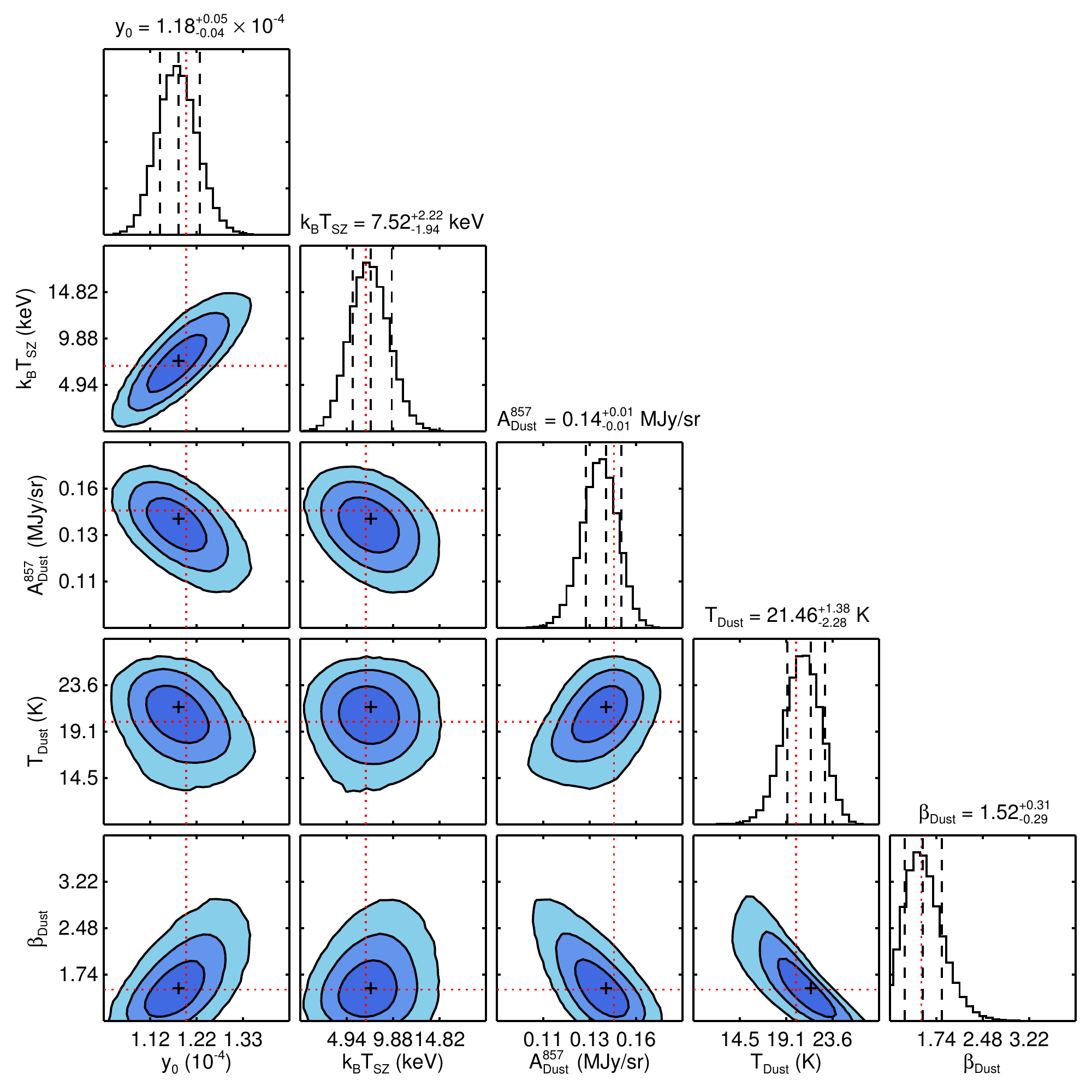}
\caption[]{Spectral modelling results obtained from simulated all-sky mock data. These results represent a single foreground realization out of 30 that are shown in Fig.~\ref{fig:bias_sz}. \textbf{Top panel:} spectrum obtained after passing the mock data through our matched filtering pipeline and stacking of the positions of the simulated clusters. The red and blue solid lines show the best-fit tSZ and FIR models, while the shaded areas indicate the envelope derived from the 68\% of models with the highest likelihood. The lower part of the panel shows the difference of the data and the best-fit model divided by the data error. As before, the shaded region corresponds to the 68\% of models with the highest likelihood. \textbf{Bottom panel:} marginalized 2D and 1D constraints on our model parameters obtained through an MCMC approach. The colours in the 2D plots represent the 68, 95, and 99\% confidence intervals. The dashed lines on top of the 1D constraints indicate the best-fit values and the 68\% 
confidence 
interval, 
while the red dotted lines indicates the true values obtained directly from the simulated tSZ and FIR maps.}
\label{fig:results_simulations}
\end{figure*}

The analysis of our mock data sets suggests that for the given subsample of the \textit{Planck~} cluster catalogue with 772 clusters the sample-average temperature can be constrained with an statistical uncertainty of $k_\mathrm{B}\Delta T_\mathrm{SZ} \approx 2 \, \mathrm{keV}$ suggesting a possible $\simeq 3\sigma$ detection, while the expected uncertainty of the Comptonization parameter corresponds to $\Delta y_0 \approx 5 \times 10^{-6}$. We furthermore find tight constraints on the parameters of our FIR model with e.g. \hbox{$A_\mathrm{Dust}^{857} = 0.14^{+0.01}_{-0.01} \, \mathrm{MJy \, sr^{-1}}$}, $T_\mathrm{Dust} = 21.5^{+1.4}_{-2.3} \, \mathrm{K}$ and $\beta_\mathrm{Dust} = 1.5^{+0.3}_{-0.3}$. We note that these low uncertainties are primarily due to the lack of a CIB model at high frequencies and that the more complex real sky will not allow for such strong constraints on the properties of cluster FIR emission. The result of excluding the CIB component in our mock data tests therefore provides somewhat 
optimistic parameter constraints but no biases.

\subsection{Simulation result: impact of the kSZ}
\label{sec:kSZ}

One of the initial assumptions in our analysis is that stacking large samples of clusters will average out the kSZ signal due to the random directions of the clusters' peculiar velocities. To test this assumption, we assign a peculiar velocity component to each of our clusters by drawing from the distribution presented by \citet{Peel06}, which is well approximated by a Gaussian with \hbox{$\sigma = 311 \, \mathrm{km \, s^{-1}}$} at $z=0$. For simplicity, we neglect the weak redshift dependence of the halo peculiar velocity and note that it will drop by about 20\% in the redshift range of $z\in[0,2]$ \citep{HM10}, making our estimates of the kSZ signal contribution a conservative one.

Using this Gaussian approximation of the velocity distribution, we can expect that after stacking the residual sample-averaged velocity should be smaller than \hbox{$311/\sqrt{772} \, \mathrm{km \, s^{-1}} \approx 11.2 \, \mathrm{km \, s^{-1}}$} with 68\% confidence. Using the optical depth of each cluster, we compute expected limits of the kSZ signal and find \hbox{$I_{217}^\mathrm{kSZ}<9.0\times10^{-4} \, \mathrm{MJy \, sr^{-1}}$} close to the peak of the kSZ spectrum, which corresponds to $0.27\sigma$ for our mock data. The situation is similar for our smaller subsample of 100 clusters with \hbox{$I_{217}^\mathrm{kSZ}<4.0\times10^{-3} \, \mathrm{MJy \, sr^{-1}}$}, corresponding to $0.3\sigma$. This demonstrates that for the given sample sizes, the kSZ can be safely neglected. At smaller sample sizes however, the kSZ can potentially lower or raise the measured intensity at $143$, $217$, and $353 \, \mathrm{GHz}$ whereas the other channels will stay mostly unaffected for all but the smallest samples.

\subsection{Simulation result: potential $Y$-bias}
\label{sec:Ybias}

 It is often assumed that relativistic corrections to the tSZ effect can be neglected. Although detecting the relativistic distortions of the tSZ spectrum for individual clusters can be beyond the reach of current experiments, ignoring the relativistic corrections can lead to a bias of the Comptonization parameter that scales with cluster temperature and therefore cluster mass. This bias will depend on the observed frequency and can be written as
\begin{equation}
 \frac{\Delta y}{y} = \frac{f(x,T_\mathrm{e})}{f(x,0 \, \mathrm{keV})} - 1.
\end{equation}
In multifrequency observations, this bias will depend on the weights assigned to each channel and thus has to be quantified through simulations. We investigate this bias using our mock data sets for two different scenarios. In the first scenario, we assume perfect foreground removal, in which case the weights for each channel will be given by the inverse squared thermal detector noise, providing the most optimistic estimate of the $y$-bias. In the second case, we clean our simulated cluster maps using an Internal Linear Combination (ILC) technique, the details of which are given in Appendix~\ref{sec:ILC}. ILC techniques are known for their robustness and simplicity and were used to produce some of the key SZ results published by the Planck Collaboration (e.g. \citealt{Planck_pressure, Planck_Coma, Planck_YMAPS}), but they require an accurate knowledge of the SZ spectral shape.
In both cases, we compute the bias on the measured cylindrically integrated Comptonization parameter within five times $R_{500}$, otherwise known as $Y_{5R500}$. Our results are summarized in Fig.~\ref{fig:y-bias}.

Our simulations show that in both cases, the integrated Comptonization parameter is systematically biased low. Fitting the measured tSZ decrement/increment signal in absence of foregrounds with a non-relativistic tSZ spectrum yields a sample-average bias of $(3\pm1)$\%. The observed bias scales roughly linear with the cluster temperature and
with $(M_{500})^{0.80\pm0.03}$. For high mass clusters, the bias can be as high as 7\% in this approach.
We find that our ILC technique produces a larger bias on average. Averaged over our entire cluster sample, the ILC estimate of the integrated $y$-parameter is biased low by $(7\pm2)$\% and up to 14\% for the hottest clusters. The bias again scales roughly linear with temperature and with $(M_{500})^{0.71\pm0.04}$ as is shown in Fig.~\ref{fig:y-bias}. The reason for this strong bias is that the ILC technique assigns a high weight to the $143 \, \mathrm{GHz}$ and $353 \, \mathrm{GHz}$ channels (see Fig.~\ref{fig:ILC_weights}) at which the difference between the relativistic and non-relativistic tSZ spectrum is particularly large. We point out that an unbiased estimation of the integrated $y$-parameter is possible by having a knowledge of the average $T_\mathrm{SZ}$ within the desired aperture while computing the ILC-weights.  More recently, \citet{Hurier17} have introduced a modified version of the ILC algorithm that is tailored to observations of the relativistic tSZ effect.

Our results demonstrate that using the non-relativistic approximation of the tSZ spectrum will lead to a systematic underestimation of the Comptonization parameter that can be as high as $14\%$ for the most massive clusters. The exact magnitude of the bias will depend on the details of the $y$-extraction method and has to be quantified and should be corrected for if possible. We further discuss this bias in Section \ref{sec:interp}.

\begin{figure}
\includegraphics{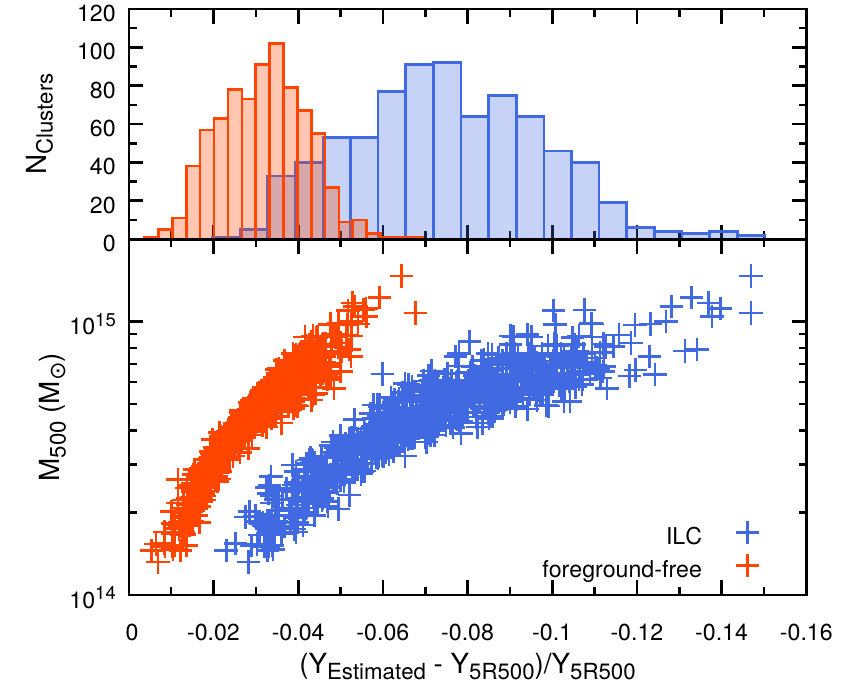}
\caption{Mass-dependent bias of the Comptonization parameter introduced by using the non-relativistic approximation of the tSZ spectrum. We quantified the bias using our mock data sets including simulated relativistic spectra. The cylindrically integrated Comptonization parameter within $5 \, \mathrm{R}_{500}$ was obtained from direct fitting of the simulated tSZ maps (red) and passing the maps though an ILC pipeline (blue, details in Appendix~\ref{sec:ILC}). In both cases, we find a mass-dependent bias of  $Y_\mathrm{5R500}$. In the former case, the values for $Y_\mathrm{5R500}$ are biased low by 3\% on average, whereas the ILC approach underestimates the true values on average by 7\%.}
\label{fig:y-bias}
\end{figure}


\section{Results from real data}
\label{sec:Results}

The main results obtained from actual \textit{Planck~}, \textit{IRAS} and \textit{AKARI} data are summarized in Fig.~\ref{fig:results_full_sample}. After matched filtering and stacking, we obtain a spectrum that clearly shows the characteristic tSZ decrement/increment plus an additional FIR excess, consistent with FIR emission from galaxy clusters. By fitting our two-component tSZ+FIR model to the extracted spectrum with $\beta_\mathrm{Dust} = 1.5$ kept fixed and marginalizing over the remaining free model parameters, we are able to constrain the average deconvolved central $y$-parameter of the sample to be $\langle y_0 \rangle = (1.24 \pm 0.04) \times 10^{-4}$, corresponding to a $31\sigma$-detection of the tSZ signal of 772 clusters. 

By modelling the relativistic distortions of the tSZ spectrum we obtain a $2.2\sigma$ measurement of the sample-average cluster temperature, which we constrain to $k_\mathrm{B}\langle T_\mathrm{SZ} \rangle = 4.4^{+2.1}_{-2.0} \, \mathrm{keV}$. Our model provides a good fit to the data with $\chi^2/\mathrm{df} = 0.98$. Furthermore, we obtain a $5\sigma$ detection of galaxy cluster-centric FIR emission with the 
FIR amplitude $A_\mathrm{Dust}^{857} = (0.10 \pm 0.2) \, \mathrm{MJy \, sr^{-1}}$. We constrain the temperature of the emitting dust grains to $T_\mathrm{Dust} = 18.4^{+3.9}_{-2.4} \, \mathrm{K}$, which is lower than the recent measurement of $(24.2\pm3.0) \, \mathrm{K}$\footnote{This value was obtained by converting the reported $(19.2\pm2.4) \, \mathrm{K}$ to the cluster rest frame using the mean redshift $\langle z \rangle = 0.26$ of the sample.} by the \citet{Planck_cluster_dust}, who performed a stacking analysis on a similar cluster sample, but with a different foreground-removal technique (see Section \ref{sec:cluster_dust}).

Due to the high uncertainties in the \textit{IRAS} and \textit{AKARI} channels, most of the constraining power comes from the \textit{Planck~} data. Excluding the \textit{IRAS} and \textit{AKARI} data points from our fit leaves the constraints on the tSZ parameters virtually untouched, while the errors on the FIR component parameters only inflate by a marginal amount to $A_\mathrm{Dust}^{857} = 0.11^{+0.02}_{-0.03} \, \mathrm{MJy \, sr^{-1}}$ and $T_\mathrm{Dust} = 18.8^{+4.0}_{-3.1} \, \mathrm{K}$ with $\chi^2/\mathrm{df} = 1.69$.

We also test for the impact of the choice of $\beta_\mathrm{Dust}$ by re-running our fit for a number of values ranging from $1.3$ up to $2.0$. We find that both  $\langle T_\mathrm{SZ} \rangle$ and $T_\mathrm{Dust}$ are anticorrelated with $\beta_\mathrm{Dust}$.
The results for spectral fitting with different fixed values for $\beta_\mathrm{Dust}$ are summarized in Table~\ref{tab:beta_table}. 

In case the SED of the cluster FIR emission varies strongly from cluster to cluster, choosing a modified blackbody as our model function can bias the tSZ parameters. We tried to account for this more complex spectrum by fitting our data with the second-order moment expansion of the modified blackbody that was introduced by \citet{Chluba12} but find that our data are not able to constrain the additional parameters related to the distribution of $T_\mathrm{Dust}$ and $\beta_\mathrm{Dust}$. We note that the distortions of the dust SED will be strongest in the Wien part at THz frequencies, where we find large errors for the \textit{IRAS} and \textit{AKARI} intensities. At \textit{Planck's} frequencies departures from the modified blackbody should be small.

In order to understand which channels have the biggest impact on the measurement of $T_\mathrm{SZ}$, we exclude individual channels one after another from the spectral fitting and record the changes of the $T_\mathrm{SZ}$ error. From this test we conclude that the \textit{Planck~} $217 \, \mathrm{GHz}$ channel is the most important one for our analysis, followed by \textit{Planck's} $545 \, \mathrm{GHz}$ channel. Excluding one of these two channels increases the uncertainty of $T_\mathrm{SZ}$ by $\simeq 50\%$, highlighting the importance of the tSZ increment for measuring temperatures via the relativistic tSZ spectrum.

In addition to our full sample of 772 clusters, we repeat our analysis for the subsample containing the 100 hottest clusters. The results of the spectral analysis of this subsample are shown in Fig.~\ref{fig:results_hot_sample}. Fitting the stacked spectrum of the clusters with the same two component tSZ+FIR model as before with \hbox{$\beta_\mathrm{Dust} = 1.5$}, we detect the tSZ signal at $23\sigma$ with \hbox{$\langle y_0 \rangle = 2.58^{+0.16}_{-0.11} \times 10^{-4}$}. We constrain the sample-average cluster temperature to $k_\mathrm{B} \langle T_\mathrm{SZ} \rangle = 6.0^{+3.8}_{-2.9} \, \mathrm{keV}$, which corresponds to a $2.0\sigma$ measurement of the tSZ relativistic corrections.
As is the case for our full sample, we observe an FIR excess at $4.4\sigma$ that is well modelled by a modified blackbody SED. For the two free parameters of the FIR model we find \hbox{$A_\mathrm{Dust}^{857} = 0.22^{+0.06}_{-0.05} \, \mathrm{MJy \, sr^{-1}}$} and \hbox{$T_\mathrm{Dust} = 16.9^{+5.0}_{-2.3} \, \mathrm{K}$}. As before, the model provides a good fit to the data with $\chi^2/\mathrm{df} = 0.69$, which changes to $\chi^2/\mathrm{df} = 1.28$ when the {\rm \textit{AKARI}} and {\rm \textit{IRAS}} data points are excluded from the fit. We note that, as for the full sample, most of the constraining power comes from the \textit{Planck~} data and excluding the additional FIR data points has little impact on our parameter constraints.

\begin{table*}
\begin{center}
\renewcommand{\arraystretch}{1.3}
\begin{tabular}{ccccccccccc}
\hline
$\beta_\mathrm{Dust}$ & & $y_0$ & $k_\mathrm{B}T_\mathrm{SZ}$ & $A_\mathrm{Dust}^{857}$ & $T_\mathrm{Dust}$ & & $y_0$ & $k_\mathrm{B}T_\mathrm{SZ}$ & $A_\mathrm{Dust}^{857}$ & $T_\mathrm{Dust}$ \\
 & &  & (keV) & ($\mathrm{MJy \, sr^{-1}}$) & (K) & &  & (keV) & ($\mathrm{MJy \, sr^{-1}}$) & (K) \\ \hline
 & | & \multicolumn{4}{{c}}{\textbf{Full sample $\bm{(n=772)}$}} & | & \multicolumn{4}{{c}}{\textbf{Hot sample $\bm{(n=100)}$}}  \\
$1.3$ & | & $1.24^{+0.04}_{-0.04} \times 10^{-4}$ & $4.38^{+2.32}_{-1.79}$ & $0.10^{+0.02}_{-0.02}$ & $21.19^{+3.62}_{-2.91}$ & | & $2.58^{+0.16}_{-0.10} \times 10^{-4}$  & $6.35^{+3.87}_{-2.97}$  & $0.22^{+0.06}_{-0.05}$  & $19.21^{+5.07}_{-2.97}$  \\ 
$1.4$ & | & $1.24^{+0.04}_{-0.04} \times 10^{-4}$ & $4.29^{+2.22}_{-1.90}$ & $0.10^{+0.02}_{-0.02}$ & $19.72^{+3.88}_{-2.57}$ & | & $2.58^{+0.17}_{-0.10} \times 10^{-4}$  & $6.39^{+3.53}_{-3.22}$  & $0.22^{+0.06}_{-0.05}$  & $18.39^{+4.62}_{-3.00}$  \\ 
$\bm{1.5}$ & | & $\bm{1.24^{+0.04}_{-0.04} \times 10^{-4}}$ & $\bm{4.36^{+2.13}_{-1.95}}$ & $\bm{0.10^{+0.02}_{-0.02}}$ & $\bm{18.44^{+3.94}_{-2.38}}$ & | & $\bm{2.58^{+0.16}_{-0.11} \times 10^{-4}}$  & $\bm{5.96^{+3.78}_{-2.93}}$  & $\bm{0.22^{+0.06}_{-0.05}}$  & $\bm{16.92^{+4.83}_{-2.26}}$  \\ 
$1.6$ & | & $1.23^{+0.05}_{-0.03} \times 10^{-4}$ & $3.98^{+2.25}_{-1.77}$ & $0.10^{+0.02}_{-0.02}$ & $17.11^{+4.05}_{-2.07}$ & | & $2.58^{+0.15}_{-0.12} \times 10^{-4}$  & $6.24^{+3.19}_{-3.45}$  & $0.21^{+0.07}_{-0.04}$  & $16.14^{+4.33}_{-2.20}$  \\ 
$1.7$ & | & $1.24^{+0.04}_{-0.04} \times 10^{-4}$ & $3.98^{+2.97}_{-1.67}$ & $0.10^{+0.02}_{-0.02}$ & $16.24^{+3.65}_{-2.02}$ & | & $2.59^{+0.14}_{-0.13} \times 10^{-4}$  & $5.68^{+3.64}_{-2.99}$  & $0.21^{+0.07}_{-0.05}$  & $15.23^{+4.36}_{-1.72}$  \\ 
$1.8$ & | & $1.23^{+0.04}_{-0.03} \times 10^{-4}$ & $4.06^{+2.18}_{-1.84}$ & $0.10^{+0.02}_{-0.02}$ & $15.54^{+3.37}_{-1.94}$ & | & $2.57^{+0.16}_{-0.11} \times 10^{-4}$  & $5.22^{+3.92}_{-2.61}$  & $0.22^{+0.07}_{-0.06}$  & $15.19^{+3.38}_{-2.21}$  \\ 
$1.9$ & | & $1.24^{+0.04}_{-0.04} \times 10^{-4}$ & $3.99^{+2.25}_{-1.77}$ & $0.10^{+0.02}_{-0.02}$ & $14.73^{+3.14}_{-1.73}$ & | & $2.56^{+0.16}_{-0.11} \times 10^{-4}$  & $4.94^{+3.98}_{-2.56}$  & $0.21^{+0.07}_{-0.05}$  & $14.61^{+3.13}_{-2.01}$  \\ 
$2.0$ & | & $1.23^{+0.05}_{-0.03} \times 10^{-4}$ & $3.92^{+2.28}_{-1.70}$ & $0.10^{+0.02}_{-0.02}$ & $14.06^{+2.80}_{-1.56}$ & | & $2.55^{+0.17}_{-0.10} \times 10^{-4}$  & $4.68^{+3.95}_{-2.47}$  & $0.21^{+0.07}_{-0.05}$  & $13.89^{+3.04}_{-1.67}$  \\ \hline
\end{tabular}
\end{center}
\caption{Parameter constraints from spectral modelling for a range of different fixed values for the spectral index $\beta_\mathrm{Dust}$ of the modified blackbody for cluster FIR emission. The constraints for $\beta_\mathrm{Dust} = 1.5$ that are highlighted in bold face are reported as our main results. We find that both $T_\mathrm{SZ}$ and $T_\mathrm{Dust}$ are anticorrelated with $\beta_\mathrm{Dust}$, whereas $\langle y_0 \rangle$ and $A_\mathrm{Dust}^{857}$ are independent of it.}
\label{tab:beta_table}
\end{table*}

\begin{figure*}
\centering

\includegraphics[width = 0.55\textwidth]{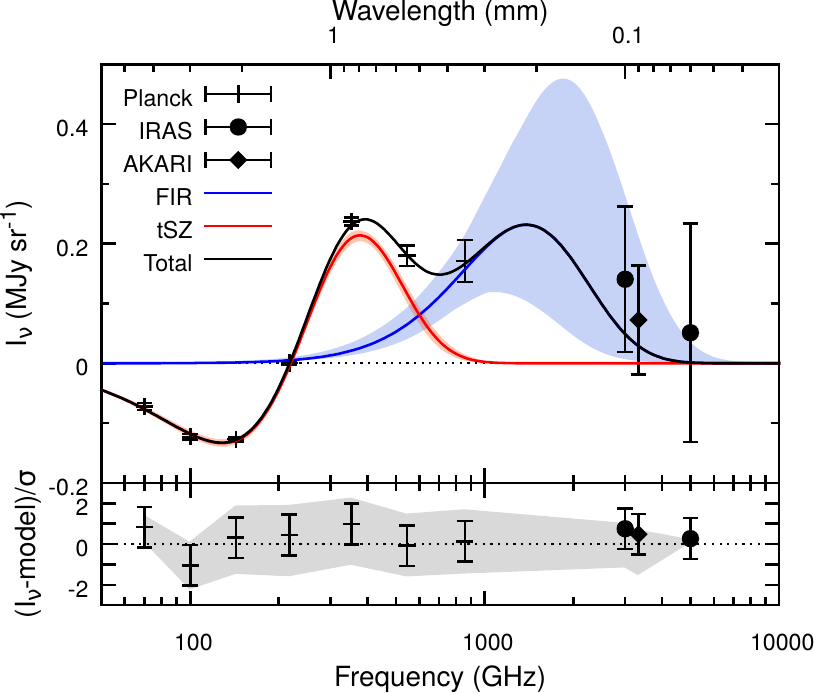}
\includegraphics[width = 0.7\textwidth]{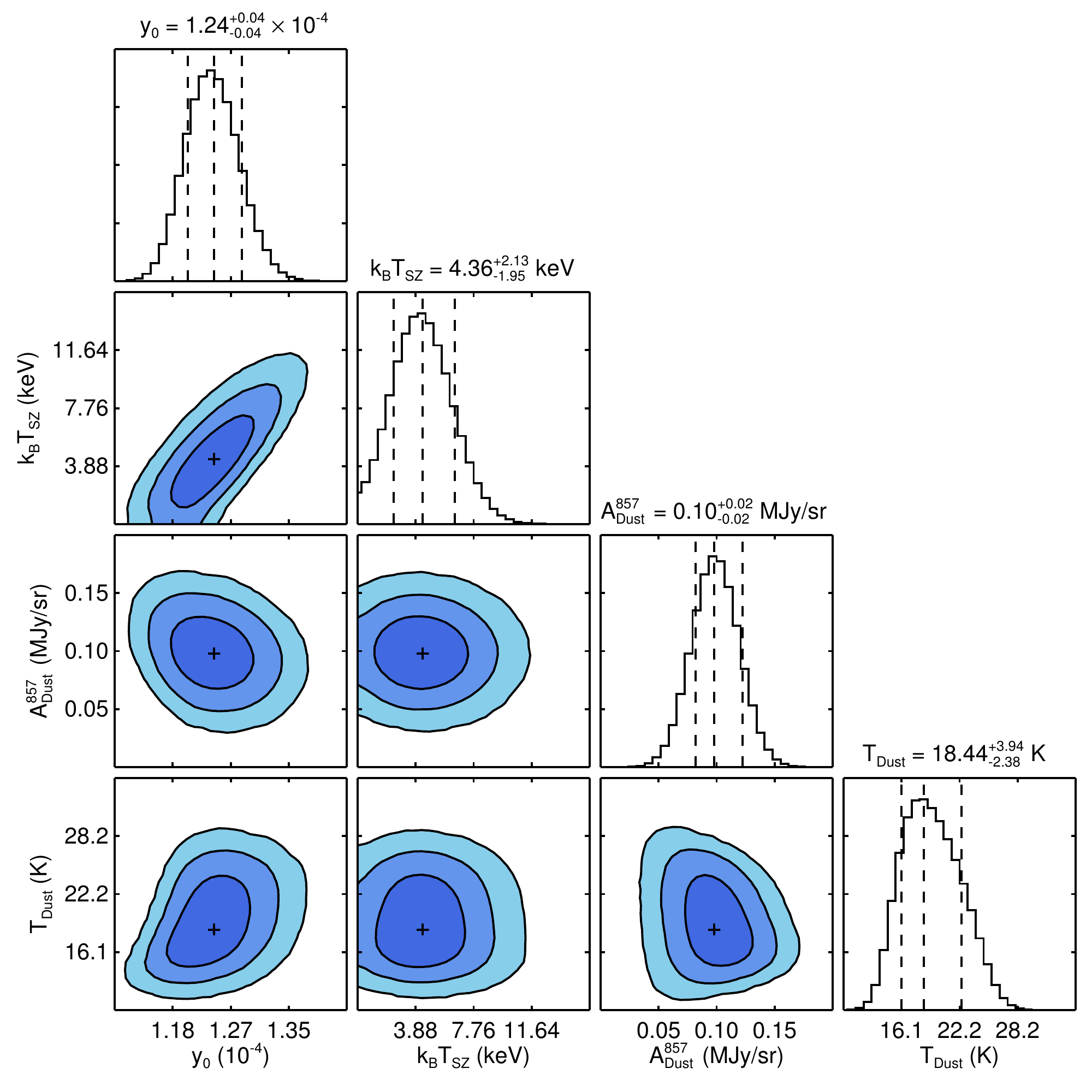}
\caption[]{Spectral modelling results for our sample of 772 galaxy clusters selected from the second \textit{Planck~} cluster catalogue (PSZ2). \textbf{Top panel:} spectrum extracted after passing the \textit{Planck~}, \textit{IRAS} and \textit{AKARI} maps through our matched filtering pipeline and stacking the cluster positions. The red and blue solid lines indicate the best-fit tSZ and FIR models. Note that the data points have been corrected for the instrumental bandpasses using the best-fit model for illustrational purposes only in order to plot smooth curves. \textbf{Bottom panel:} marginalized 2D and 1D constraints on our model parameters obtained through an MCMC approach. The colours in the 2D plots represent the 68, 95, and 99\% confidence intervals. The dashed lines on top of the 1D constraints indicate the best-fit values and the 68\% confidence interval. The third parameter of the FIR model $\beta_\mathrm{Dust}$ was fixed to the common value 1.5 in order to obtain these results. We do not observe any strong correlation between 
the tSZ and 
FIR parameters. The tSZ signal of the sample is detected with high significance ($31\sigma$) and we obtain a $2.2\sigma$ measurement of the sample-average cluster temperature.}
\label{fig:results_full_sample}
\end{figure*}

\begin{figure*}
\centering

\includegraphics[width = 0.55\textwidth]{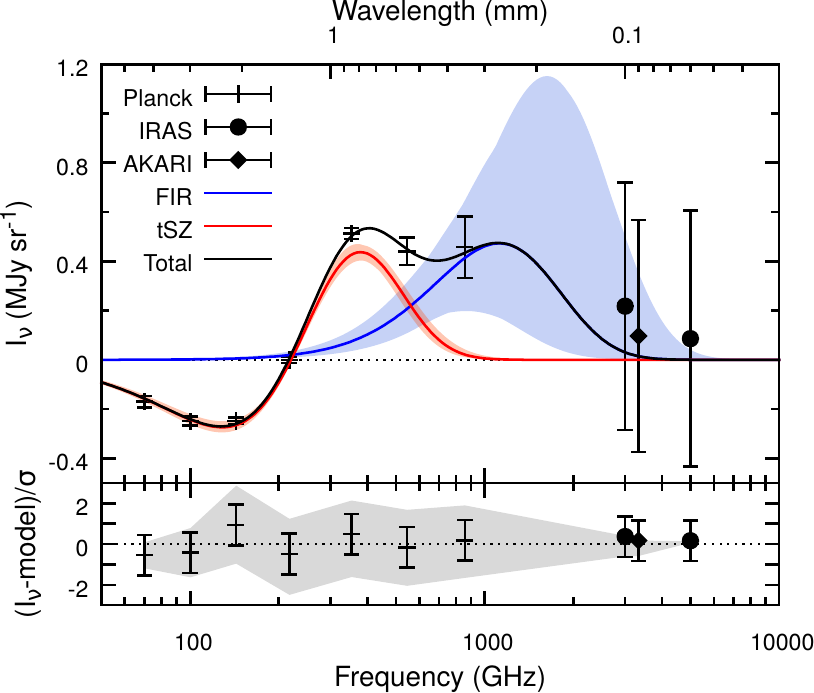}
\includegraphics[width = 0.7\textwidth]{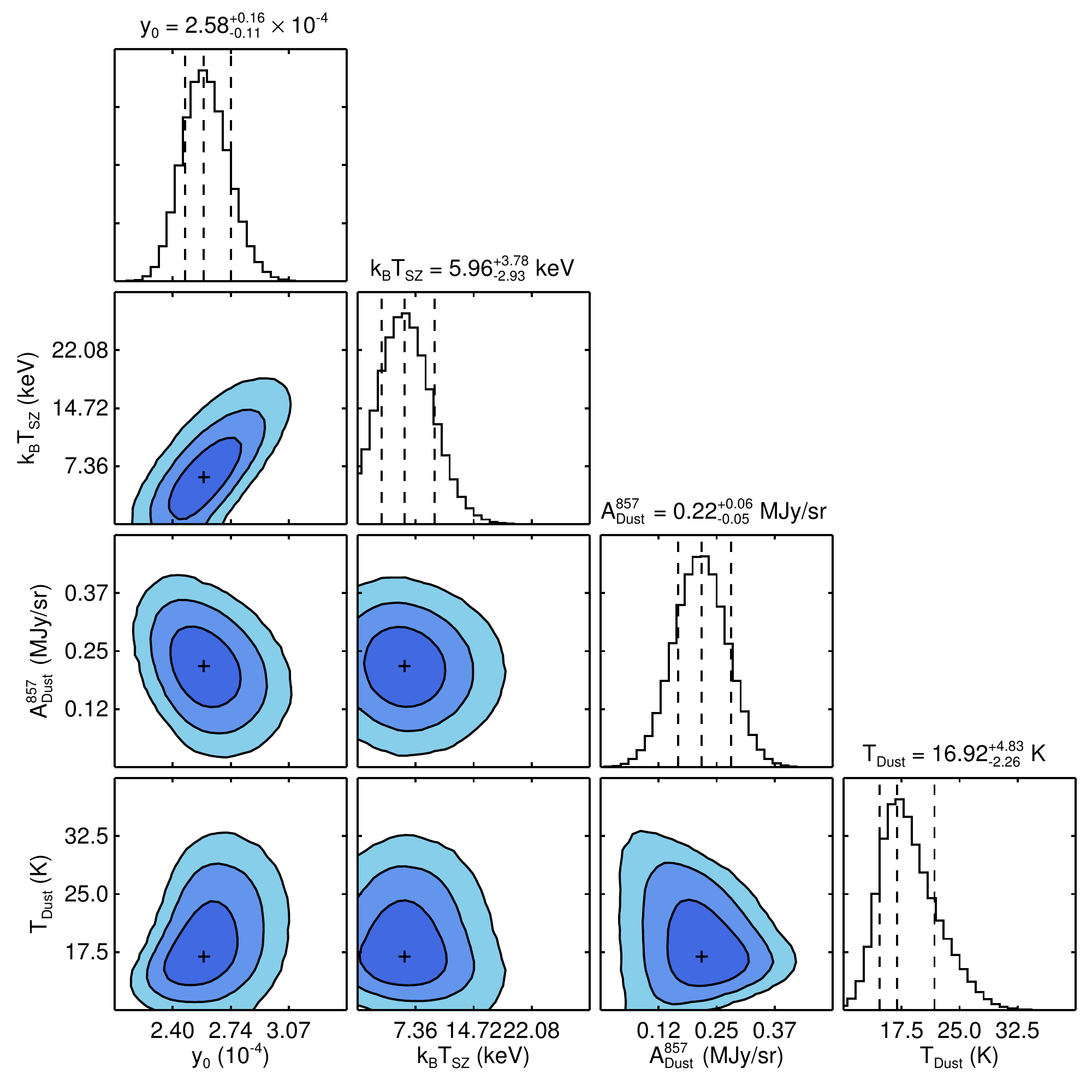}
\caption[]{Spectral modelling results for a smaller subsample, containing just the 100 hottest clusters as determined through the $M$--$T$ scaling relation given by equation~(\ref{eq:M-T_relation}). \textbf{Top panel:} as before, we show the spectrum extracted after passing the \textit{Planck~}, \textit{IRAS} and \textit{AKARI} maps through our matched filtering pipeline and stacking the cluster positions. The red and blue solid lines indicate the best-fit tSZ and FIR models. \textbf{Bottom panel:}  marginalized 2D and 1D constraints on our model parameters obtained through an MCMC approach for the 100 hottest clusters. As for the full sample, the third parameter of the FIR model $\beta_\mathrm{Dust}$ was fixed to $1.5$. We again do not observe a strong correlation of the tSZ and FIR model parameters. Although the average $y$-parameter of the clusters is roughly twice as high as for the full sample, the significance of the tSZ signal detection reduces to $23\sigma$. We are able to measure a higher sample-average cluster temperature, 
consistent with 
our expectation, but at a slightly reduced significance of $2.0\sigma$.}
\label{fig:results_hot_sample}
\end{figure*}


\section{Discussion}
\label{sec:Discussion}

\subsection{Interpretation of the main results}
\label{sec:interp}

After careful signal extraction and spectral fitting, we can confirm the signature of a relativistic tSZ (or rSZ) signal in \textit{Planck~} full-mission data at roughly 95\% significance level. For our sample of 772 clusters, we find an average temperature of \hbox{$k_\mathrm{B} \langle T_\mathrm{SZ} \rangle = 4.4^{+2.1}_{-2.0} \, \mathrm{keV}$}, which is consistent with the mass-weighted average X-ray temperature $k_\mathrm{B} \langle T_\mathrm{X} \rangle = (6.91 \pm 0.07) \, \mathrm{keV}$.
There is a tentative difference (at roughly $1.3\sigma$) between these two values, with $\langle T_\mathrm{SZ} \rangle$ being lower than the sample-averaged X-ray spectroscopic temperature $\langle  T_\mathrm{X} \rangle$.

We find that, due to the sensitivity of \textit{Planck~}, a better constraint on the relativistic tSZ-derived temperature is not obtainable by simply selecting the hottest clusters from the cluster catalogue. While this approach increases the mean sample temperature, the noise also increases due to the smaller sample size. As a result, the detection significance of $\langle T_\mathrm{SZ}\rangle$ remains roughly constant. The best-fit value of $k_\mathrm{B} \langle T_\mathrm{SZ} \rangle = 6.0^{+3.8}_{-2.9} \, \mathrm{keV}$ in this subsample is again lower but consistent with the expected mass-weighted X-ray temperature $k_\mathrm{B} \langle T_\mathrm{X} \rangle = (8.54 \pm 0.16) \, \mathrm{keV}$.

The lower $T_\mathrm{SZ}$ values can result from the different weighting schemes in tSZ and X-ray temperature measurements. While $T_\mathrm{SZ}$ is weighted linearly with the gas density, $T_\mathrm{X}$ is weighted with its square. Previous studies showed that the gas mass-weighted temperature $T_\mathrm{mg}$, which behaves similar to $T_\mathrm{SZ}$, measured within an aperture to be lower than the X-ray spectroscopic temperature \citep{Vikhlinin06, Nagai07} and the ratio $Y_\mathrm{SZ}/Y_\mathrm{X}$ to be less than unity \citep{Arnaud10}. We investigate the impact of the weighting schemes in Appendix \ref{sec:temperatures} using analytical temperature and density profiles and find that $T_\mathrm{X}$ is higher than $T_\mathrm{SZ}$ by $\sim 20\%$ for non-cool-core clusters when averaged within $\theta_{500}$ and only lower then $T_\mathrm{SZ}$ for cool-core clusters when small apertures ($\lesssim 0.3 \theta_{500}$) are used. 
We note that our cluster sample is a representative subset of the Planck PSZ2 clusters, since the sample selection is only affected by Galactic foregrounds and point sources. Recently, \citet{Rossetti17} found the cool-core fraction of a representative subset of \textit{Planck~} clusters to be $\approx30\%$. Therefore we do not expect the ratio $T_\mathrm{X}/T_\mathrm{SZ}$ observed within dense cool cores to significantly affect our results.
On the other hand, hydrodynamic simulations frequently produce a large number of cold and dense clumps that are able to bias $T_\mathrm{X}$ low compared to $T_\mathrm{SZ}$ (or $T_\mathrm{mg}$) within the entire cluster volume (e.g. \citealt{Kay08, Biffi14}), yet more recent and improved simulation codes predict the dissociation of such clumps \citep{Beck16}.  It is beyond the scope of the current paper to make a detailed analysis of this ratio that will require a systematic evaluation of the $T_\mathrm{X}$  measurements in the parent samples of \citet{Reichert11} from which our $T_\mathrm{X}$  scaling is taken, for example whether spectral fits were obtained after masking dense substructures within the clusters or not.

Even though \textit{Planck~} data do not provide evidence for the relativistic distortions in the tSZ spectrum with high significance, the presence of these distortions can nevertheless cause a bias in the measured SZ signals when non-relativistic spectra are used to extract Comptonization $y$-maps or fit data in a matched multifiltering approach. We demonstrated this bias in Section \ref{sec:Ybias} through our simulated mock cluster sample with realistic noise and foregrounds. A similar analysis based on the application of ILC algorithms on simulated maps for the Cosmic ORigins Explorer (CORE) mission has been presented by \citet{Hurier17}, who find bias up to 20\% in the $Y$-value of the hottest clusters.

The bias lowers the measured $Y$-value and is mass-dependent. A mass dependence is expected since the relativistic corrections to the spectrum would only be significant for high-mass clusters. It is interesting to note that the direction and mass dependence of this bias are both similar to the so-called hydrostatic mass bias that is assumed in the cosmological analysis of \textit{Planck~} clusters. This bias term, parameterized  by a $(1-b)$ factor in $Y$--$M$ scaling relations (e.g., \citealt{PlanckClusterCosmo13}), accounts for all possible biases in the mass measurement and the use of the non-relativistic spectrum for the tSZ signal extraction will certainly be a part of it. As we do not follow the exact SZ signal extraction methods (matched multifiltering and {\small POWELLSNAKES}) that are used by the Planck Collaboration and also do not carry out the steps necessary to connect $Y_{500}$ to $M_{500}$ via X-ray mass proxies, 
we are unable to comment on the exact bias on the $Y_{500}$--$M_{500}$ scaling relation used in the \textit{Planck~} analysis. 

We instead focus on quantifying the $Y$-measurement bias based on our mock data, finding it to be around 5\% (optimistic case with no foregrounds) up to about 14\% (extreme case based on the ILC method) for the most massive clusters.}
The mass dependence of the $Y$-bias is also of interest, which we found to be approximately $(M_{500})^{0.71\pm 0.04}$ when using the ILC approach. This is very similar to the slope of the hydrostatic mass bias found in weak-lensing mass calibration of subsets of \textit{Planck~} clusters, for example by \citet{vdL14}, who found a mass scaling between the \textit{Planck~} SZ and weak-lensing mass estimates having a power law index of $0.68^{+0.15}_{-0.11}$. Even though it is expected that more massive systems would show stronger deviations from hydrostatic equilibrium due to their enhanced mass accretion rate (e.g., \citealt{Shi14}), the similar mass dependence of both these biases suggests that the observed effect can be a combination of the two.

We also consider the effect of electron temperature variance within our mock cluster sample. As explained in \citet{Chluba13}, the second moment of the temperature field causes another correction to the average SZ signal. Using our mock data, we find the $y$-weighted temperature moments, $k_\mathrm{B}\langle T_{\rm e} \rangle_y \simeq 7.7\,{\rm keV}$ and $k_\mathrm{B}\langle T_{\rm e}^2 \rangle_y \simeq 64\,{\rm keV^2}$, implying $\sigma_{T_{\rm e}}\simeq 2.1\,{\rm keV}$. At the current level of sensitivity, this leads to a negligible correction to the sample-averaged relativistic SZ signal and can be ignored. However, future precision measurements of $T_\mathrm{SZ}$ in multiple mass bins might offer a possibility to constrain the slope of the cluster mass function using higher-order moments of $T_\mathrm{e}$.

\subsection{Comparison with other works}
\label{sec:hurier}

Recently, \citet{Hurier16} claimed the first high significance detection of the tSZ relativistic corrections by stacking \textit{Planck~} maps of clusters taken from the \hbox{X-ray}-selected MCXC cluster catalogue \citep{Piffaretti11}, as well as several smaller cluster catalogues with \hbox{X-ray} spectroscopic temperatures. \citet{Hurier16} binned clusters from both the MCXC cluster catalogue and the combined spectroscopic catalogue according to their temperature. A comparison of the obtained tSZ inferred ICM temperatures $T_\mathrm{SZ}$ with $T_\mathrm{X}$ revealed a linear trend with a significance of $3.7\sigma$ for the MCXC clusters and $5.3\sigma$ for the spectroscopic ones, which is the main result reported by the author. In addition \citet{Hurier16} finds that the $T_\mathrm{SZ}$ values are higher than $T_\mathrm{X}$ with a ratio $T_\mathrm{SZ}/T_\mathrm{X}$ of $1.65\pm0.45$ and $1.38\pm0.26$, respectively.

The approach presented in this work differs from the one used by \citet{Hurier16} mostly in the foreground removal and spectral modelling techniques. \citet{Hurier16} adopted the foreground removal approach presented by \citet{Hurier14}, in which Galactic and extragalactic thermal dust emission is subtracted by using the \textit{Planck~} $857 \, \mathrm{GHz}$ channel as a template that is extrapolated to lower frequencies using a scale factor. This scale factor is computed for each channel under the assumption that the SED is constant in a $2^\circ \times 2^\circ$ field around each cluster, excluding the central 30$\,$arcmin. The \textit{Planck~} $217 \, \mathrm{GHz}$ channel is used analogously to remove the contribution of the CMB from the remaining maps, making use of its well-known frequency spectrum.

We note that subtracting the $217 \, \mathrm{GHz}$ and $857 \, \mathrm{GHz}$ maps to remove the CMB and Galactic dust can lead to a distortion of the tSZ spectrum of the clusters due to the non-negligible tSZ signal within these two \textit{Planck~} bands. This can be understood using the tabulated, band-integrated spectra provided in Table~\ref{tab:SZspec}. Assuming a typical dust SED with $T_\mathrm{Dust} = 20 \, \mathrm{K}$ and $\beta = 1.5$, the intensity at $545 \, \mathrm{GHz}$ is approximately $33\%$ of the intensity at $857 \, \mathrm{GHz}$ and $9\%$ in case of $353 \, \mathrm{GHz}$. For a $10 \, \mathrm{keV}$ cluster subtracting the $857 \, \mathrm{GHz}$ map thus reduces the tSZ signal at $545 \, \mathrm{GHz}$ by about $47 \, \mathrm{MJy \, sr^{-1}}$ to $907 \, \mathrm{MJy \, sr^{-1}}$, corresponding to $T_\mathrm{e} \approx 5 \, \mathrm{keV}$. Analogous calculations can be done to estimate the impact of subtracting the $217 \, \mathrm{GHz}$ map and show that the bias will be largest for low-temperature 
systems. In 
addition to the partial subtraction of the tSZ signal, the assumption of a constant dust SED across the field neglects the redshift-induced \textit{K}-correction needed for the cluster FIR emission.

Our work relies on a matched filtering approach to reduce the Galactic and extragalactic foregrounds, which only assumes that these are spatially uncorrelated with the cluster signal. The correlated cluster FIR component is accounted for later in our spectral modelling. The validity of our approach is tested with mock data. Clusters with known low-frequency point sources are removed from our sample and we therefore do not include a dedicated low-frequency component in our spectral model. 

Our attempts at constraining the $T_\mathrm{X}$ versus $T_\mathrm{SZ}$ linear slope using a cluster sample similar to the one used by \citet{Hurier16} with direct X-ray spectroscopic temperatures produce inconclusive results. Starting from the same cluster catalogues with spectroscopic $T_\mathrm{X}$ information, we obtain a total of 313 clusters after removal of duplicates and applying our Galactic mask and point source flagging. This sample, when split into three temperature bins, yields large errors that leave $T_\mathrm{SZ}$ unconstrained. We repeat this analysis with our full sample of 772 clusters, with $T_\mathrm{X}$ values estimated using equation~(\ref{eq:M-T_relation}) and grouped into four temperature bins. Fixing the line intercept at $[0 \, \mathrm{keV},0 \, \mathrm{keV}]$, we find the normalization of the ratio $T_\mathrm{SZ}/T_\mathrm{X}$ to be smaller than unity at roughly $2 \sigma$ significance. This result and its errors are similar to the values derived earlier for our full sample and the 
subsample 
containing the 100 hottest clusters.

\subsection{FIR emission from galaxy clusters}
\label{sec:cluster_dust}

In recent years, it has been shown that clusters are sources of FIR emission. Although the exact nature of this emission remains uncertain, current observations point towards both dusty cluster members, as well as stripped warm dust in the ICM. Furthermore, clusters act as powerful gravitational lenses of the CIB, the magnified emission of which further adds to the observed emission.
This spatially correlated FIR emission makes accurate measurements of the relativistic tSZ more challenging and requires joint spectral modelling of both components.
Matched filtering techniques like the one employed by us are particularly suited to separate the FIR emission from clusters from Galactic and uncorrelated CIB emission with similar SED based on their spatial distribution.

\begin{figure}
\includegraphics{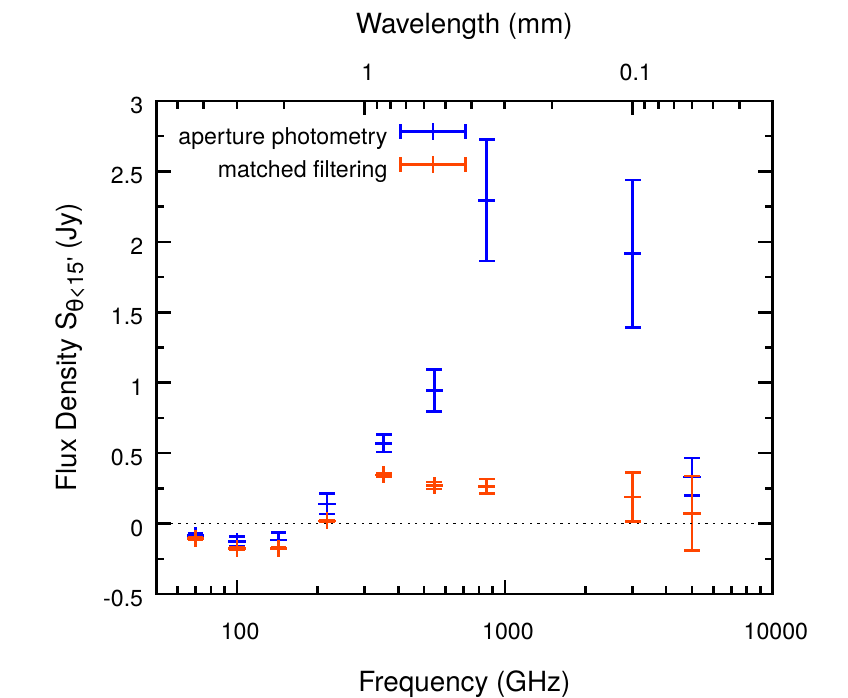}
\caption{Comparison of the fluxes extracted with our matched filtering technique (red) and the aperture photometry method (blue). The latter has been used in previous studies conducted by the Planck Collaboration (\citeyear{Planck_tSZE_CIB, Planck_cluster_dust}). We use both techniques to extract the stacked spectrum of our sample of 772 clusters integrated within 15$\,$arcmin. We find that matched filtering offers much cleaner maps with less Galactic residuals, as reflected by the smaller errorbars, and reduces the contribution from cluster FIR emission to the extracted spectrum significantly.}
\label{fig:aperture_photometry}
\end{figure}

To demonstrate this, we compare our method against the frequently used `aperture photometry' method of foreground removal, and the results are shown in Fig.~\ref{fig:aperture_photometry}. The matched filters are constructed and applied in the same way as described in Section \ref{sec:matched_filtering}, with the exception that we compute the signal integrated within 15$\,$arcmin, which is achieved by multiplying the deconvolved amplitude that is returned by the filter with the integral of the cluster profile:
\begin{equation}
 S_\nu (<r) = I_\nu^\mathrm{filt} 2\pi \int_0^{\theta'} \theta y(\theta) \, \mathrm{d}\theta, 
\end{equation}
where $I_\nu^\mathrm{filt}$ is the stacked flux after filtering. In case of the aperture photometry technique, we integrate the signal in the stacked maps within the same 15$\,$arcmin aperture and subtract the background that is constrained from an annulus with $15\arcmin<r<60\arcmin$. The errorbars are derived by performing the same steps on 1000 randomly positioned stacked fields. Our comparison shows that matched filtering allows for less Galactic foreground residuals resulting in smaller errorbars and reduces the contribution of cluster FIR emission to the observed signal significantly. Spectral fitting of the spectrum obtained through aperture photometry delivers a higher dust temperature \hbox{$T_\mathrm{Dust} = 22.6^{+0.9}_{-1.3} \, \mathrm{K}$} compared to matched filtering. This is closer to the value reported by the \citet{Planck_cluster_dust}. Even though we measure the dust temperature at higher significance compared to the values reported in Section \ref{sec:Results} 
due 
to the increased FIR amplitude, the larger errors at low frequencies do not allow to constrain the average electron temperature of the clusters.

Although \textit{Planck's} resolution does not allow us to determine the exact nature of the FIR emission from clusters, we can explore its scaling with cluster mass and redshift. Due to the redshift-dependent selection of the \textit{Planck~} clusters, splitting the entire sample into two mass or redshift  bins will produce correlated results; hence, we restrict these variables for the following analysis.
We find that half of our sample (i.e. 386 clusters) lies within a relatively narrow redshift interval $0.072 < z < 0.257$, allowing us to minimize a potential redshift evolution of the dust luminosity. We split this sample into a low-mass and a high-mass subsample with 193 clusters each (\hbox{$\langle M_{500}^\mathrm{high} \rangle = 5.1\times10^{14} \, \mathrm{M_\odot}$}, \hbox{$\langle  M_{500}^\mathrm{low} \rangle = 3.0\times10^{14} \, \mathrm{M_\odot}$}, \hbox{$\langle z^\mathrm{high} \rangle = 0.192$}, \hbox{$\langle  z^\mathrm{low} \rangle = 0.132$}). After fixing the SED of the FIR component by assuming $T_\mathrm{Dust} = 20 \, \mathrm{K}$ for both samples, the observed dust amplitudes and average sample masses are assumed to be related via a power law: 
\begin{equation}
 \frac{A_\mathrm{Dust}^\mathrm{high}}{A_\mathrm{Dust}^\mathrm{low}} = \left(\frac{1+\langle z^\mathrm{low}\rangle}{1+\langle z^\mathrm{high}\rangle} \right)^4 \left( \frac{\langle M_{500}^\mathrm{high} \rangle}{\langle M_{500}^\mathrm{low} \rangle} \right)^\epsilon.
\end{equation}
We find that the observed dust amplitude scales with the cluster mass to the power of $\epsilon = 0.8^{+1.7}_{-1.3}$. This value is consistent with the value of 1.0 that is assumed by the \citet{Planck_tSZE_CIB}, but is significantly smaller than the value of  $4.4\pm1.0$ that can be derived from the dust masses reported by the \citet{Planck_cluster_dust} for two different cluster mass bins. 
Our analysis is limited by the large uncertainties on $A_\mathrm{Dust}$ (\hbox{$A_\mathrm{Dust}^\mathrm{high} = (0.07 \pm 0.03) \, \mathrm{MJy \, sr^{-1}}$}, \hbox{$A_\mathrm{Dust}^\mathrm{low} = 0.04^{+0.03}_{-0.02} \, \mathrm{MJy \, sr^{-1}}$}).

We investigate the redshift evolution of the FIR emission by repeating this test analogously for a low-$z$ and a high-$z$ subsample. We find that half of our sample lies within the cluster mass interval $ 3.7 \times 10^{14} \, \mathrm{M_\odot} < M_{500} < 6.0 \times 10^{14} \, \mathrm{M_\odot}$, which we than split into a low-redshift subsample with $\langle z^\mathrm{low} \rangle = 0.191$ and a high redshift subsample with $\langle z^\mathrm{high} \rangle = 0.387$. We again use a power law to relate the observed FIR amplitudes to the redshifts of the subsamples
\begin{equation}
 \frac{A_\mathrm{Dust}^\mathrm{high}}{A_\mathrm{Dust}^\mathrm{low}} = \left(\frac{1+\langle z^\mathrm{high}\rangle}{1+\langle z^\mathrm{low}\rangle} \right)^{\delta-4},
\end{equation}
and constrain the power law slope of the redshift dependence to be $\delta = 6\pm3$ with \hbox{$A_\mathrm{Dust}^\mathrm{high} = 0.13^{+0.05}_{-0.04} \, \mathrm{MJy \, sr^{-1}}$} and \hbox{$A_\mathrm{Dust}^\mathrm{low} = 0.09^{+0.04}_{-0.03} \, \mathrm{MJy \, sr^{-1}}$}.

Detailed studies of the mass and redshift dependence of the infrared luminosity and the related dust content of clusters will be exciting goals for the next generation of sub-mm/FIR observatories. With the increased sensitivities that will be provided by future observatories, the assumption of a single temperature modified blackbody is likely to break down. As a consequence, more complex models that account for a temperature variance along the line of sight like the one presented by \citet{Chluba17} might be needed.

\subsection{Outlook: CCAT-prime}

In the final section, we discuss what future SZ experiments might be able to improve upon the constraints on the relativistic tSZ-derived temperature. Currently, \textit{Planck~} is the only experiment with the necessary spectral coverage to model the entire tSZ/relativistic tSZ spectrum and separate its contribution from cluster FIR emission. Future space-based experiments, similar to the CORE mission \citep{Delab17}, will have the same spectral coverage as \textit{Planck~}, but with many more spectral channels and far better sensitivity making them ideally suited for this kind of measurements. 
In addition, future CMB spectrometers, similar to the Primordial Inflation  Explorer (PIXIE; \citealt{PIXIE2011}), would improve upon the sensitivity of \textit{COBE} FIRAS experiment by several orders of magnitude and 
are expected to detect the average relativistic thermal SZ at very high significance ($\simeq 10-20\sigma$, \citealt{Hill15, Abitbol2017}), although their angular resolution may not allow for a study of individual clusters.  
Ground-based experiments proposed under the CMB-S4 concept\footnote{\url{https://cmb-s4.org}} will have a restricted frequency range, capped at around $270 \, \mathrm{GHz}$, but more than two orders of magnitude better sensitivity than \textit{Planck~} that will also enable a detailed modelling of the SZ spectrum. Here we present result predictions for a new telescope, named CCAT-prime, that is expected to start its observation well ahead of these two other classes of experiments and can provide relativistic tSZ-based temperature measurements of individual clusters.

CCAT-prime (CCAT-p for short) will be a $6 \, \mathrm{m}$ diameter submillimetre telescope operating at 5600 m altitude in the Chilean Atacama desert.
The high and dry site on a mountaintop in the Chajnantor plateau will offer excellent atmospheric conditions for submillimetre continuum surveys up to 350 $\mu$m wavelength \citep{Bustos14}, and the high-throughput optical design will allow for large focal-plane arrays similar to the future CMB-S4 experiments \citep{Niem16}. Beginning its first-light observations in 2021, CCAT-p will perform large area multiband surveys for the SZ effect. We consider the sensitivities for a fiducial 4000$\,$h, 1000$\,$deg$^2$ survey in seven frequency bands with an instrument based on the design presented by \cite{Stacey14}. The expected survey sensitivities are quoted in Table~\ref{tab:CCATp} in comparison to the \textit{Planck~} full-mission data. It is seen that the individual channel sensitivities for CCAT-p are about a factor of $\simeq 5-15$ better, except for the highest frequency band, for which the sky emissivity is roughly 40\% from the ground even in the best quartile of weather. 

 \begin{table}
  \begin{center}
  \tabcolsep=0.13cm
  \begin{tabular}{ccccc}
  \hline
  $\nu$ & FWHM & $\Delta T$  & $\Delta T$ & $\Delta I$ \\
  (GHz) & (arcmin) & ($\mathrm{\mu K_{RJ}}$-arcmin) & ($\mathrm{\mu K_{CMB}}$-arcmin) & (kJy/sr-arcmin) \\ \hline
  \multicolumn{5}{{c}}{\textbf{\textit{Planck} (all-sky-average full mission data)}}  \\
  100 & 9.68 & 61.4 & 77.3 & 18.9 \\
  143 & 7.30 & 19.8 & 33.4 & 12.4 \\
  217 & 5.02 & 15.5 & 46.5 & 22.5 \\
  353 & 4.94 & 11.7 & 156 & 44.9 \\
  545 & 4.83 & 5.1 & 806 & 46.8 \\
  857 & 4.64 & 1.90 & $1.92\times10^4$ & 43.5 \\ \\
  \multicolumn{5}{{c}}{\textbf{CCAT-p (4000$\,$h, 1000$\,$ deg$^2$ survey)}}  \\
  95 & 2.2 & 3.9 & 4.9 & 1.1 \\
  150 & 1.4 & 3.7 & 6.4 & 2.6 \\
  226 & 1.0 & 1.5 & 4.9 & 2.4 \\
  273 & 0.8 & 1.2 & 6.2 & 2.7 \\
  350 & 0.6 & 2.0 & 25 & 7.6 \\
  405 & 0.5 & 2.9 & 72 & 15 \\
  862 & 0.3 & 3.9 &  $6.6\times10^4$ & 89 \\ \hline
  \end{tabular}
  \caption{Comparison of the noise characteristics and spatial resolution of CCAT-p and \textit{Planck}. The values for \textit{Planck~} represent all-sky averages \citep{Planck_overview} that were scaled to arcmin scale under the assumption of white noise. The values for CCAT-p are representative of a 4000$\,$h, 1000 deg$^2$ survey performed under average weather condition. The intrinsic beam sensitivities are again given at arcmin scale.}
  \label{tab:CCATp}
  \end{center}
  \end{table} 
  
\begin{figure*}
\centering
\includegraphics[width = 0.7\textwidth]{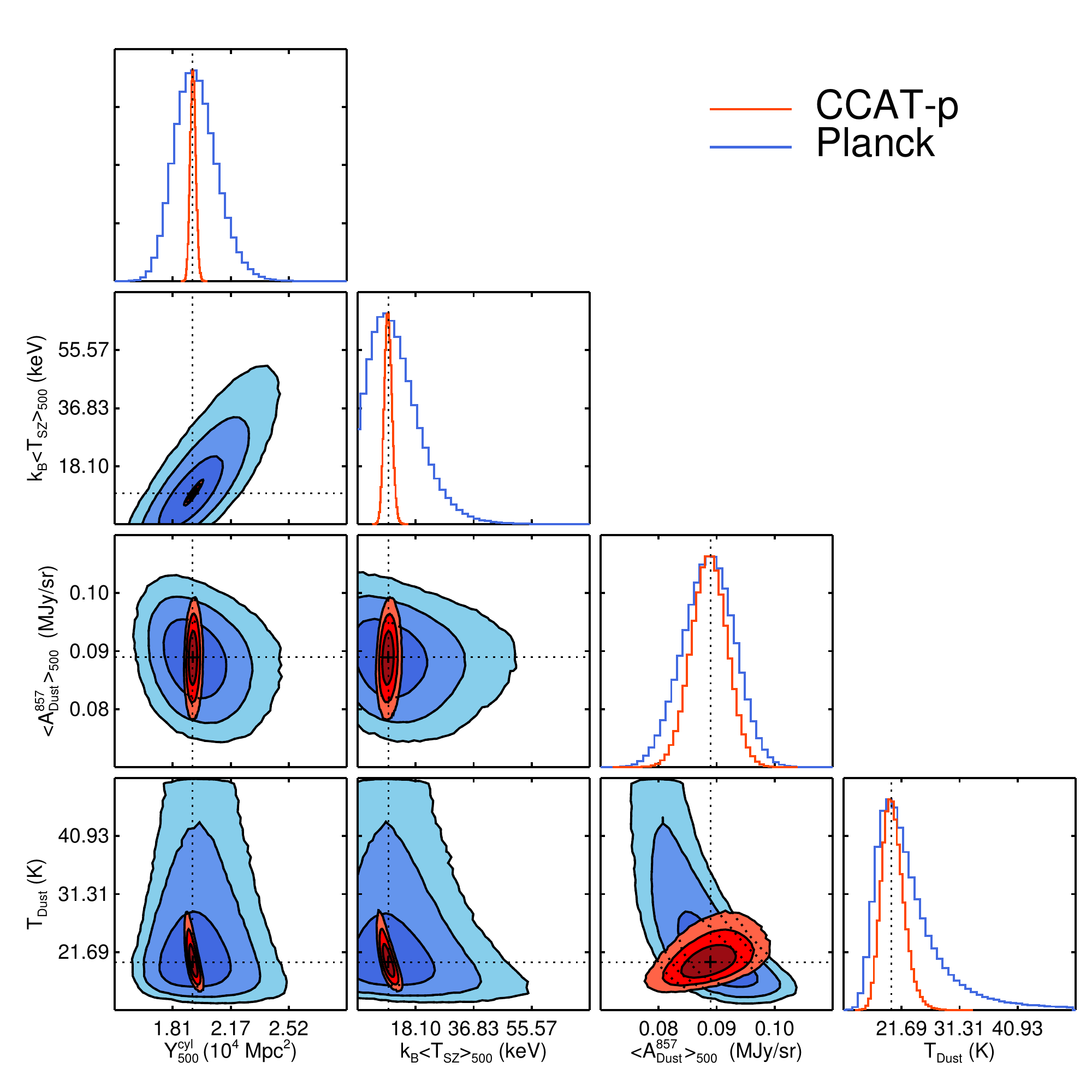}
\caption[]{Spectral modelling comparison between the \textit{Planck~} and CCAT-prime sensitivities. The blue contours correspond to \textit{Planck~} and the red ones to \hbox{CCAT-p}, while the dotted lines indicate the input values. 
We neglect the foreground emissions and the kSZ signal arising from the cluster peculiar motion, considering only the thermal noise as listed in Table~\ref{tab:CCATp}. The modelling results are for a single $M_{500} = 10^{15}$ $\mathrm{M}_{\odot}$ cluster at $z=0.23$ with dust mass $M_\mathrm{Dust} = 5\times 10^{10} \, \mathrm{M_\odot}$. The improved sensitivity of the CCAT-p in the $95$--$405 \, \mathrm{GHz}$ range helps to place far better constraints on the cluster SZ parameters than \textit{Planck~}, while having roughly the same constraining power on the cluster dust temperature and amplitude as in \textit{Planck~} due to the lower sensitivity in the $862 \, \mathrm{GHz}$ channel. The temperature in this single high-mass cluster is constrained at $9 \sigma$ for the CCAT-p survey sensitivity.}
\label{fig:CCATp}
\end{figure*}

We carry out a simplified comparison between \textit{Planck~} and CCAT-p for constraining the cluster SZ and dust parameters that ignores all Galactic and extragalactic foregrounds (thus also not taking advantage of the roughly six times better angular resolution compared to \textit{Planck~} for matched filtering). We consider a high mass \hbox{($M_{500} = 10^{15}$ $\mathrm{M}_{\odot}$)} cluster at $z=0.23$ with a dust mass of $5 \times 10^{10} \, \mathrm{M_\odot}$ and $T_\mathrm{Dust} = 20 \, \mathrm{K}$, which we simulate using the same model that was introduced in Section \ref{sec:simulation_set-up}. The results of our analysis are summarized in Fig.~\ref{fig:CCATp}. Thanks to the roughly one order of magnitude better sensitivity in the $95$--$405 \, \mathrm{GHz}$ frequency range, the CCAT-p survey will be able to determine the temperature of this single high-mass cluster with high precision from the survey data (\hbox{CCAT-p}: \hbox{$Y_{500}^\mathrm{cyl} = 1.93^{+0.02}_{-0.01} \times 10^{-4} \, \mathrm{Mpc^2}$}, 
\hbox{$k_\mathrm{B} T_{\mathrm{SZ}} = 9.1^{+1.5}_{-1.0} \, \mathrm{keV}$}; \textit{Planck~}: \hbox{$Y_{500}^\mathrm{cyl} = 1.92^{+0.15}_{-0.10}\times 10^{-4} \, \mathrm{Mpc^2}$}, \hbox{$k_\mathrm{B} T_{\mathrm{SZ}} = 9.3^{+10.6}_{-4.5} \, \mathrm{keV}$}). 
The cluster FIR emission is constrained roughly at the same level of precision as with \textit{Planck~} data, although the better angular resolution ($0.2 \ \mathrm{arcmin}$ at $862 \, \mathrm{GHz}$) will help for more accurate point source removal (\hbox{CCAT-p}: \hbox{$A_\mathrm{Dust}^{857} = (88\pm4) \, \mathrm{kJy \, sr^{-1}}$}, \hbox{$T_\mathrm{Dust} = 20.0^{+2.2}_{-1.4} \, \mathrm{K}$}; \textit{Planck~}: \hbox{$A_\mathrm{Dust}^{857} = (88\pm6) \, \mathrm{kJy \, sr^{-1}}$}, \hbox{$T_\mathrm{Dust} = 20.0^{+5.5}_{-2.6} \, \mathrm{K}$}). 
By excluding individual channels from the spectral fitting, we infer that the $405 \, \mathrm{GHz}$ has the biggest impact on the constrain on $T_\mathrm{SZ}$ for the CCAT-p survey, while the $862 \, \mathrm{GHz}$ channel is crucial for measuring the properties of the FIR component.
  
 We find that when including a cluster velocity component ($\varv_\mathrm{pec} = 500 \, \mathrm{km \, s^{-1}}$) and fitting simultaneously for the kSZ signal, the uncertainty of the tSZ parameters $Y_{500}^\mathrm{cyl}$ and $T_\mathrm{SZ}$ increases by roughly 50\%, while the peculiar velocity is constraint to $\varv_\mathrm{pec} = 521^{+76}_{-62} \, \mathrm{km \, s^{-1}}$. The SZ and dust parameters show very little correlation, resulting in almost unchanged constraints on the dust amplitude and temperature. In contrast, when adding a kSZ component we are neither able to constrain the peculiar velocity nor the dust temperature from our simulated \textit{Planck~} data without assigning strong priors. Further predictions for kSZ observations of clusters with CCAT-p are given by \citet{Mittal18}.
 
We note again that the limits quoted here are only for an idealized comparison between the two instruments when foregrounds are neglected. Our results nevertheless highlight the potential of the upcoming CCAT-p telescope to radically improve on \textit{Planck~} and push the limits of ground based observations. The performance of CCAT-p will be explored under more realistic circumstances in forthcoming papers.

  
\section{Conclusions}
\label{sec:Conclusion}

The tSZ effect has become a widely used tool for finding mass-selected cluster samples, since its signal is proportional to the thermal energy of the intracluster medium and thus to the total cluster mass. 
In addition to the integrated pressure, the spectrum of the tSZ effect also contains information on the ICM temperature as the thermal electrons with keV energies inside massive galaxy clusters will distort the tSZ spectrum towards higher frequencies, resulting in an effect that is second-order in cluster temperature.  
These relativistic corrections to the tSZ effect are commonly accounted for when modelling the tSZ signal from massive clusters, but only recently have direct measurements become feasible. In this paper, we present the first attempt to constrain the relativistic signal in the tSZ spectrum by directly modelling it together with cluster FIR emission within a wide frequency range.

The detection of the relativistic tSZ effect requires to measure both the decrement and increment of the tSZ spectrum using a set of massive galaxy clusters. Both these requirements are satisfied by the data from the \textit{Planck~} mission, which is the only current data set that has the necessary spectral coverage and sensitivity and also provides an almost complete catalogue of the most massive clusters in the observable universe. We set out to stack the multifrequency data of a well-selected sample of 772 galaxy clusters. Our modelling includes an FIR component associated with galaxy clusters, which has been established by several recent measurements, and we employ data from the \textit{IRAS} and \textit{AKARI} missions in addition to the \textit{Planck~} HFI channels to augment the stacked tSZ$+$FIR spectrum at THz frequencies.

One important aspect of our analysis is realistic simulations with mock clusters that are used for the validation of our approach. We show that the cluster model parameters recovered through our spectral fitting method are free from any significant biases and that the kSZ signal from the cluster peculiar motions are effectively averaged out by stacking a large number of clusters. With these simulated data sets, we also show that measuring the integrated Compton $y$-parameter by using a non-relativistic spectrum, as is done for \textit{Planck~} and other SZ survey data, can result in a non-negligible bias towards lower $Y$-values. For the most massive clusters in the \textit{Planck~} catalogue, we compute this bias to be around $5$--$14$\%, depending on the method. This bias also carries a moderate mass dependence that scales (in the units of $Y_\mathrm{5R500}$) approximately as $(M_{500})^{0.7-0.8}$. 

Results from stacking the all-sky data from \textit{Planck~} provide significant, but not fully conclusive evidence for the relativistic tSZ signal. When stacking our full sample of 772 clusters, we are able to measure the tSZ relativistic corrections at $2.2\sigma$, constraining the mean temperature of this sample to be $4.4^{+2.1}_{-2.0} \, \mathrm{keV}$. We repeat the same analysis on a subsample containing only the 100 hottest clusters, for which we measure the mean temperature to be $6.0^{+3.8}_{-2.9} \, \mathrm{keV}$, corresponding to $2.0\sigma$. In contrast to some recently published results, we find that these average $T_\mathrm{SZ}$ values appear to be lower than the corresponding $\langle T_\mathrm{X}\rangle$ values.
This might be a systematic trend due to the different weighting schemes of SZ and X-ray temperature measurements, which lead to $T_\mathrm{X}>T_\mathrm{SZ}$ averaged within $\theta_{500}$ and beyond if gas clumping is moderate. However, the large uncertainties of our $T_\mathrm{SZ}$ measurements do not permit a more detailed analysis.

In our analysis, the temperature of the emitting dust grains that cause the observed cluster FIR emission is constrained to $\simeq 20 \, \mathrm{K}$, consistent with previous studies. The measured amplitude of our FIR model is roughly one order of magnitude lower than those reported in earlier works, which used aperture photometry for signal extraction.
This demonstrates the superiority of the matched filtering technique in removing all spatially uncorrelated foregrounds as well as reducing any cluster specific emission with a spatial distribution that differs from the SZ signal. 
We probe the mass and redshift dependence of the cluster FIR signal amplitude and find that with the current data we cannot constrain a power law mass dependence, although there is some evidence for strong scaling of the cluster FIR emission with redshift.

As a final outlook we provide predictions for a future ground-based submillimetre survey experiment called CCAT-prime. Using the sensitivity estimates of a fiducial CCAT-prime survey of 4000$\,$h, we find that this experiment will be able to constrain the cluster SZ parameters with roughly $5$--$10$ times higher precision than \textit{Planck~}, therefore being able to determine the temperature of individual high-mass systems using the relativistic tSZ signal. Similarly improved constraints (compared to \textit{Planck~} data) are obtained when a kSZ signal due to the peculiar motion of clusters is added to the model. 
Such high-precision data will bring a new era of SZ measurements of galaxy clusters in which the relativistic tSZ effect can be used to obtain an independent measurement of the ICM temperature, thereby breaking the degeneracy between the density and temperature from tSZ measurements, providing a more complete thermodynamical description of the intracluster medium from SZ data alone.  

\section*{Acknowledgements}
\addcontentsline{toc}{section}{Acknowledgements}
We would like to thank the referee for providing several insightful suggestions that helped to improve the paper. We also thank Dominique Eckert, Benjamin Magnelli, Jean-Baptiste Melin, Bjoern Soergel, Ricardo G{\'e}nova-Santos, and Mark Vogelsberger for useful discussions. The authors acknowledge Gordon Stacey and Michael Niemack for providing the CCAT-prime survey characteristics and Miriam Ramos Ceja, Ana Mikler, and Katharina Bey for helpful comments on the draft for this manuscript.
We acknowledge the use of the Centre d'Analyse de Donn\'{e}es Etendues (CADE).
J.E.  acknowledges  support  by  the  Bonn-Cologne  Graduate  School  of  Physics and  Astronomy  (BCGS).
J.E., K.B., and F.B. acknowledge partial funding from the Transregio programme TRR33 of the Deutsche Forschungsgemeinschaft (DFG).
J.C. is supported by the Royal Society as a Royal Society University Research Fellow at the University of Manchester, UK.
%

\appendix

\section{Tabulated SZ spectra}
 Table~\ref{tab:SZspec} provides the tSZ spectrum including relativistic corrections from $0 \, \mathrm{keV}$ to $20 \, \mathrm{keV}$ computed with {\small SZPACK} \citep{Chluba12} for the \textit{Planck~} $70$--$857 \, \mathrm{GHz}$ channels. The spectra include corrections for the \textit{Planck~} instrumental bandpass that were computed as presented by the Planck Collaboration (\citeyear{Planck_HFI}). Assuming $y=1$, we provide the spectra both in units of specific intensity ($\mathrm{MJy \, sr^{-1}}$) as well as $\mathrm{K_{CMB}}$. In the former case, we provide the intensity decrement/increment \hbox{$\Delta I_\mathrm{tSZ}(x, T_\mathrm{e}) = y I_0 \, h(x) \, f(x, T_\mathrm{e})$} as given by equation~(\ref{eq:bandpass}). In units of $\mathrm{K_{CMB}}$, we provide \hbox{$\Delta T_\mathrm{tSZ}(x, T_\mathrm{e}) = y \, T_\mathrm{CMB} \, f(x, T_\mathrm{e})$} for which we find
\begin{equation}
    \Delta\tilde{T}_\mathrm{tSZ}(x, T_\mathrm{e}) = y \, T_\mathrm{CMB} \, \frac{ \int \mathrm{d}\nu \, \tau(\nu) \, h(x)\, f_\mathrm{rel}(x, T_\mathrm{e})}{\int \mathrm{d}\nu \, \tau(\nu) \, h(x)}.
\end{equation}

\begin{table*}
\begin{center}
\tabcolsep=0.13cm
\begin{tabular}{ccccccccccccccc}
\hline
 $k_\mathrm{B}T_\mathrm{e}$ & \multicolumn{2}{c}{$70 \, \mathrm{GHz}$} & \multicolumn{2}{c}{$100 \, \mathrm{GHz}$}  & \multicolumn{2}{c}{$143 \, \mathrm{GHz}$}  & \multicolumn{2}{c}{$217 \, \mathrm{GHz}$}  & \multicolumn{2}{c}{$353 \, \mathrm{GHz}$} & \multicolumn{2}{c}{$545 \, \mathrm{GHz}$}  & \multicolumn{2}{c}{$857 \, \mathrm{GHz}$}  \\
 (keV) & (MJy/sr) & ($\mathrm{K_{CMB}}$) & (MJy/sr) & ($\mathrm{K_{CMB}}$)  & (MJy/sr)  & ($\mathrm{K_{CMB}}$)  & (MJy/sr)  & ($\mathrm{K_{CMB}}$)  & (MJy/sr)  & ($\mathrm{K_{CMB}}$)  & (MJy/sr)  & ($\mathrm{K_{CMB}}$)  & (MJy/sr)  & ($\mathrm{K_{CMB}}$)  \\ \hline
0 & -637.9 & -4.938 & -981.6 & -4.021 & -1034.7 & -2.784 & 93.5 & 0.193 & 1784.4 & 6.207 & 838.7 & 14.451 & 59.7 & 26.323 \\ 
1 & -634.9 & -4.915 & -969.7 & -3.973 & -1027.9 & -2.765 & 83.9 & 0.173 & 1758.4 & 6.117 & 855.8 & 14.746 & 66.9 & 29.502 \\ 
2 & -631.7 & -4.889 & -963.5 & -3.947 & -1020.7 & -2.746 & 72.9 & 0.151 & 1732.4 & 6.027 & 870.9 & 15.007 & 74.5 & 32.836 \\ 
3 & -628.4 & -4.864 & -957.3 & -3.922 & -1013.7 & -2.727 & 62.3 & 0.129 & 1707.0 & 5.938 & 884.9 & 15.247 & 82.4 & 36.318 \\ 
4 & -625.2 & -4.839 & -951.2 & -3.897 & -1006.8 & -2.708 & 52.0 & 0.107 & 1682.2 & 5.852 & 897.7 & 15.468 & 90.6 & 39.925 \\ 
5 & -622.0 & -4.815 & -945.2 & -3.872 & -1000.2 & -2.691 & 42.0 & 0.087 & 1658.0 & 5.768 & 909.4 & 15.670 & 99.0 & 43.636 \\ 
6 & -618.9 & -4.790 & -939.3 & -3.848 & -993.6 & -2.673 & 32.3 & 0.067 & 1634.4 & 5.686 & 920.2 & 15.856 & 107.6 & 47.428 \\ 
7 & -615.7 & -4.766 & -933.5 & -3.824 & -987.3 & -2.656 & 22.9 & 0.047 & 1611.3 & 5.605 & 930.0 & 16.025 & 116.3 & 51.283 \\ 
8 & -612.6 & -4.742 & -927.8 & -3.801 & -981.0 & -2.639 & 13.9 & 0.029 & 1588.6 & 5.527 & 939.0 & 16.179 & 125.2 & 55.186 \\ 
9 & -609.5 & -4.718 & -922.1 & -3.778 & -975.0 & -2.623 & 5.1 & 0.010 & 1566.5 & 5.450 & 947.1 & 16.320 & 134.1 & 59.120 \\ 
10 & -606.5 & -4.695 & -916.6 & -3.755 & -969.0 & -2.607 & -3.4 & -0.007 & 1544.9 & 5.374 & 954.5 & 16.447 & 143.1 & 63.070 \\ 
11 & -603.5 & -4.671 & -911.1 & -3.732 & -963.2 & -2.591 & -11.7 & -0.024 & 1523.7 & 5.301 & 961.2 & 16.562 & 152.0 & 67.027 \\ 
12 & -600.5 & -4.648 & -905.7 & -3.710 & -957.5 & -2.576 & -19.7 & -0.041 & 1502.9 & 5.228 & 967.2 & 16.665 & 161.0 & 70.977 \\ 
13 & -597.5 & -4.625 & -900.3 & -3.688 & -951.9 & -2.561 & -27.5 & -0.057 & 1482.6 & 5.158 & 972.6 & 16.758 & 169.9 & 74.912 \\ 
14 & -594.6 & -4.603 & -895.1 & -3.667 & -946.5 & -2.546 & -35.0 & -0.072 & 1462.7 & 5.089 & 977.3 & 16.840 & 178.8 & 78.822 \\ 
15 & -591.7 & -4.580 & -889.9 & -3.646 & -941.1 & -2.532 & -42.3 & -0.088 & 1443.2 & 5.021 & 981.6 & 16.913 & 187.6 & 82.701 \\ 
16 & -588.8 & -4.558 & -884.8 & -3.625 & -935.8 & -2.517 & -49.4 & -0.102 & 1424.1 & 4.954 & 985.3 & 16.977 & 196.3 & 86.542 \\ 
17 & -585.9 & -4.536 & -879.7 & -3.604 & -930.7 & -2.504 & -56.3 & -0.116 & 1405.4 & 4.889 & 988.5 & 17.033 & 204.9 & 90.337 \\ 
18 & -583.1 & -4.514 & -874.7 & -3.584 & -925.6 & -2.490 & -63.0 & -0.130 & 1387.0 & 4.825 & 991.3 & 17.080 & 213.4 & 94.084 \\ 
19 & -580.3 & -4.492 & -869.8 & -3.563 & -920.6 & -2.477 & -69.5 & -0.144 & 1369.0 & 4.763 & 993.6 & 17.121 & 221.8 & 97.777 \\ 
20 & -577.5 & -4.471 & -865.0 & -3.544 & -915.7 & -2.463 & -75.8 & -0.157 & 1351.4 & 4.701 & 995.5 & 17.154 & 230.0 & 101.412 \\ \hline
\end{tabular}
\end{center}
\caption{Tabulated, bandpass integrated tSZ spectra with relativistic corrections computed with {\small SZPACK} for $y=1$. The spectra are provided for all \textit{Planck~} channels used in the main analysis and for electron temperatures ranging from $0 \, \mathrm{keV}$ (non-relativistic) up to $20 \, \mathrm{keV}$. The tSZ is negligible in the \textit{IRAS} and \textit{AKARI} bands at thermal temperatures. We provide the spectra in units of both $\mathrm{K_{CMB}}$ and specific intensity. Please note that we compute the tSZ spectrum on a much finer temperature grid for the main analysis and allow temperatures up to $75 \, \mathrm{keV}$.}
\label{tab:SZspec}
\end{table*}

\section{Null tests}

Our analysis follows the approach of \citet{Soergel17} who use a similar stacking approach of \textit{Planck~}, \textit{IRAS}, and \textit{AKARI} data to search for active galactic nucleus feedback in quasi-stellar objects (QSOs) with the tSZ effect. As part of their results, the authors reported a statistically significant offset of $-1.5 \, \mathrm{mJy}$ in their matched-filtered IRIS $100 \, \micron$ map after stacking random positions as part of a null test. This result is obtained by employing a similar matched filtering approach to the one used in this work. As possible reasons for this offset, \citet{Soergel17} name striping errors or calibration uncertainties and excluded the $100 \, \micron$ channel from their main analysis.

\begin{figure}
\includegraphics{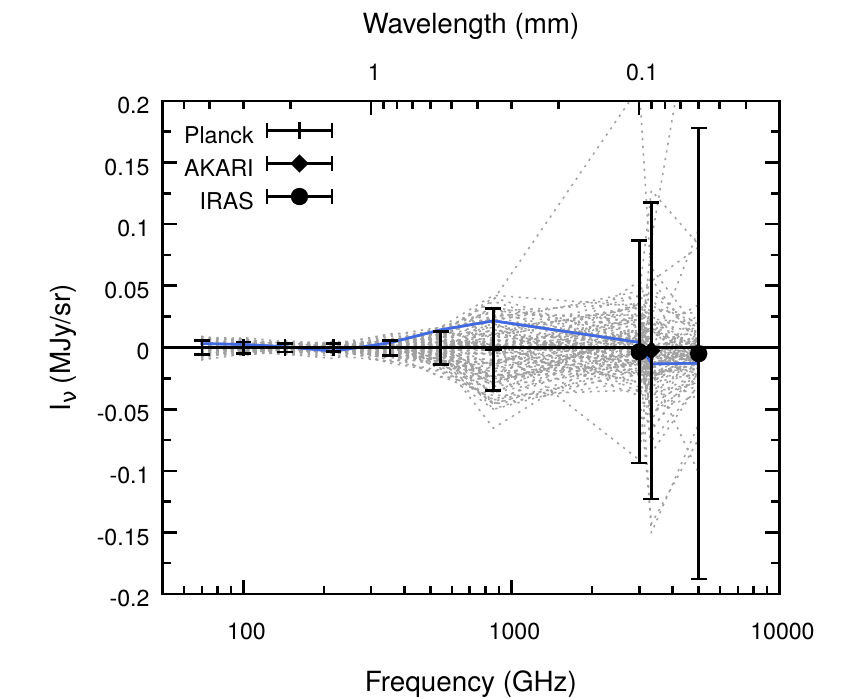}
\caption{Results of a null test performed by stacking random positions after matched filtering of each map. Each realization, shown as grey dotted lines, was produced by stacking 772 positions, equal to the number of cluster in our sample. The black data points indicate the average of all realizations, while the blue solid line highlights a single representative realization.}
\label{fig:null_test}
\end{figure}

We conduct a similar test by stacking $772$ random positions uniformly sampled across the sphere outside the same 40\% Galactic mask used for sample selection. This step is repeated $10 \, 000$ times to produce a sufficiently high number of realizations to obtain an estimate of the channel-to-channel covariance matrix used in our main analysis and double as the data for our null test. The result of this test is shown in Fig.~\ref{fig:null_test}. 

Our results demonstrate that none of the instruments shows a significant offset in any of the used channels. We find that the average signal at $100 \, \micron$ is consistent with $0$ and we are unable to reach the precision necessary to test the findings reported by \citet{Soergel17}. The different results of the null tests are likely due to the non-uniform sampling of the random positions mimicking the distribution of QSOs based on Sloan Digital Sky Survey data as adopted by \citet{Soergel17}. 
The authors also employ a much larger sample size of $377 \, 136$ QSOs and optimize their filters to recover point-like sources, whereas we optimize our filters for galaxy clusters. To test whether a larger sample size reveals a weak bias, we stack $\sim 300 \, 000$ random positions to roughly match the QSO sample size and find an average intensity of $(-0.0001 \pm 0.0036) \, \mathrm{MJy \, sr^{-1}}$ in the stacked $100 \, \micron$ map. Our results thus indicate that, after matched filtering, the \textit{Planck~} $70$--$857 \, \mathrm{GHz}$, \textit{IRAS} $100$ and $60 \, \micron$ as well as the \textit{AKARI} $90 \, \micron$ channel do not show a statistically significant bias within the boundary of our method.

\section{The ILC technique}
\label{sec:ILC}

The ILC algorithm \citep{Bennett03} is a popular technique for the removal of foregrounds in multifrequency CMB observations. It is a so-called semiblind approach to foreground removal; that is, it only requires precise knowledge of the frequency spectrum of the desired astrophysical signal making it an ideal tool for the extraction of CMB maps. While doing so, no prior information or auxiliary data from other observations is needed, which is the reason for the term `internal'. The method makes two key assumptions: 
\begin{enumerate}
 \item The observed maps are a linear mixture of astrophysical components and instrumental noise.
 \item The individual components are uncorrelated.
\end{enumerate}
Using the first assumption and following the approach presented by \citet{Hurier13}, the $N_\nu$ observed maps $\bm{I}(p)$ can be written as
\begin{equation}
 \mathbfit{I}(p) = \mathbfss{A} \bm{S}(p) + \bm{N}(p),
\end{equation}
where $p$ denotes the map pixels, $\mathbfss{A}$ is the mixing matrix that contains the spectral information
of the $N_s$ astrophysical components and has the dimensions $N_\nu \times N_s$, $\bm{S}(p)$ is a vector that contains the $N_s$ astrophysical components and $\bm{N}(p)$ is the vector containing the instrumental noise of the $N_\nu$ channels.

The ILC technique assumes that an estimate of an astrophysical component of interest can be obtained by forming a linear combination of the observed maps
\begin{equation}
 \hat{S}_\mathrm{ILC}(p) = \bm{\omega}^\mathrm{T} \bm{I}(p) = \sum_{i=1}^{N_\nu} = \omega_i I_i(p).
\end{equation}
Following \citet{Eriksen04}, the variance of the map  $\hat{\mathbfss{S}}_\mathrm{ILC}$ can be written as
\begin{equation}
 Var(\hat{\mathbfss{S}}_\mathrm{ILC}) = \bm{\omega}^\mathrm{T} \hat{\mathbfss{C}} \bm{\omega},
\end{equation}
where $\hat{\mathbfss{C}}$ is the empirical covariance matrix of the observed maps
\begin{equation}
  \hat{C}_{ij} = COV(I_i, I_j) \equiv \frac{1}{N_\mathrm{pix}} \sum^{N_\mathrm{pix}}_{p=1} (I_i(p)-\langle I_i \rangle)(I_j(p)-\langle I_j \rangle).
  \label{eqn:covariance}
\end{equation}
The ILC-weights $\bm{\omega}^\mathrm{T}$ are determined by minimizing the variance of $\hat{\mathbfss{S}}_\mathrm{ILC}$
\begin{equation}
  \frac{\partial}{\partial \, \omega_{\mathrm{i}}}\left[\bm{\omega}^\mathrm{T} \hat{\mathbfss{C}} \bm{\omega}\right]=0.
\end{equation}
In addition, the weights $\bm{\omega}$ are required to have unit response to the component of interest in order to preserve its signal, i.e.
\begin{equation}
 \bm{\omega}^\mathrm{T} \bm{a}  = 1,
\end{equation}
where $\bm{a}$ is the mixing vector of the component of interest. In case of the tSZ, the mixing vector will be $\bm{a} = \Delta I_\mathrm{tSZ} / y = I_0 h(\bm{x}) f(\bm{x}, T_\mathrm{e})$. We note that instrument-specific bandpass corrections like presented in equation~(\ref{eq:bandpass}) will have to be applied.
Furthermore, \citet{Remazeilles11_CILC} showed that additional astrophysical components with well-known frequency spectra $\bm{b_i}$ can be removed in a constrained ILC approach by demanding 
\begin{equation}
 \bm{\omega}^\mathrm{T} \bm{b}_i  = 0.
\end{equation}
We combine the mixing vector of the component of interest together with the mixing vectors of all $N$ constrained unwanted components into the matrix $\mathbfss{F}$ of dimensions $N_\nu \times (1+N)$
\begin{equation}
  \mathbfss{F}=\begin{pmatrix}
    a[1]  & b_1[1] & \dots & b_N[1]  \\
    \vdots      & \vdots       & \ddots& \vdots \\
    a[N_\nu] & b_1[N_\nu] & \dots & b_N[N_\nu] \\
    \end{pmatrix}.
\end{equation}
A solution to this optimization problem can be found by solving a linear system using Lagrange multipliers $\bm{\lambda}$ 
\begin{equation}
   \begin{pmatrix}
    2\cdot \hat{\mathbfss{C}}  & - \mathbfss{F} \\
    \mathbfss{F}^\mathrm{T} & 0
   \end{pmatrix} 
   \begin{pmatrix}
    \bm{\omega} \\
    \bm{\lambda} 
   \end{pmatrix}
   =
   \begin{pmatrix}
    0 \\
    \bm{e} 
   \end{pmatrix},
 \end{equation}
where $\bm{e} = (1,0,...,0)^\mathrm{T}$ is the $(1+N)$ vector containing the response of the constrained astrophysical components to the ILC-weights. The solution to this problem is given by 
\begin{equation}
  \bm{\omega}^\mathrm{T} = \bm{e}^\mathrm{T}\left(\mathbfss{F}^\mathrm{T} \hat{\mathbfss{C}}^{-1}\mathbfss{F} \right)^{-1} \mathbfss{F}^\mathrm{T} \hat{\mathbfss{C}}^{-1}.
\end{equation}

Since the ILC technique requires precise knowledge of the frequency spectrum of the component of interest, any deviation from the correct spectral shape will lead to a bias in the estimate of the component map $\hat{\mathbfss{S}}_\mathrm{ILC}$. This is particularly problematic because the covariance matrix of the observed maps and thus the ILC-weights are usually computed over a large field of several square degree. Therefore, even when relativistic corrections are included, a bias will be present in most pixels of the map because galaxy clusters are not isothermal and a single large field will contain multiple clusters. If a non-relativistic tSZ spectrum is used to compute the ILC-weights, the $y$-bias will be given by
\begin{equation}
 \frac{\Delta y}{y} = \bm{\omega}^\mathrm{T} \bm{a}_\mathrm{tSZ}^\mathrm{rel} -1,
\end{equation}
where $\bm{a}_\mathrm{tSZ}^\mathrm{rel}$ is the relativistic tSZ mixing vector for a given temperature. 

In order to investigate the bias caused by using the non-relativistic tSZ spectrum, we use mock data sets created using the steps presented in Section \ref{sec:simulations}. We compute the covariance matrix of the simulated maps from $10^\circ \times 10^\circ$ cut-outs around our simulated clusters and constrain the CMB spectrum, which in units of specific intensity will have the mixing vector $\bm{b}_\mathrm{CMB} = h(\bm{x})$. For simplicity, we do not employ spatial decomposition techniques like the ones that are used by the {\small NILC} \citep{Remazeilles11_NILC} and {\small MILCA} \citep{Hurier13} algorithms. We then apply the obtained ILC-weights directly to the simulated SZ decrement/increment maps in order to obtain estimates of the $y$-maps that are unaffected by foreground residuals and instrumental noise, but are determined using realistic data. The results are shown in the main text.
We verify our algorithm by also simulating mock data sets featuring non-relativistic clusters, in which case, as expected, we do not observe any bias in $y$.

\begin{figure}
\includegraphics{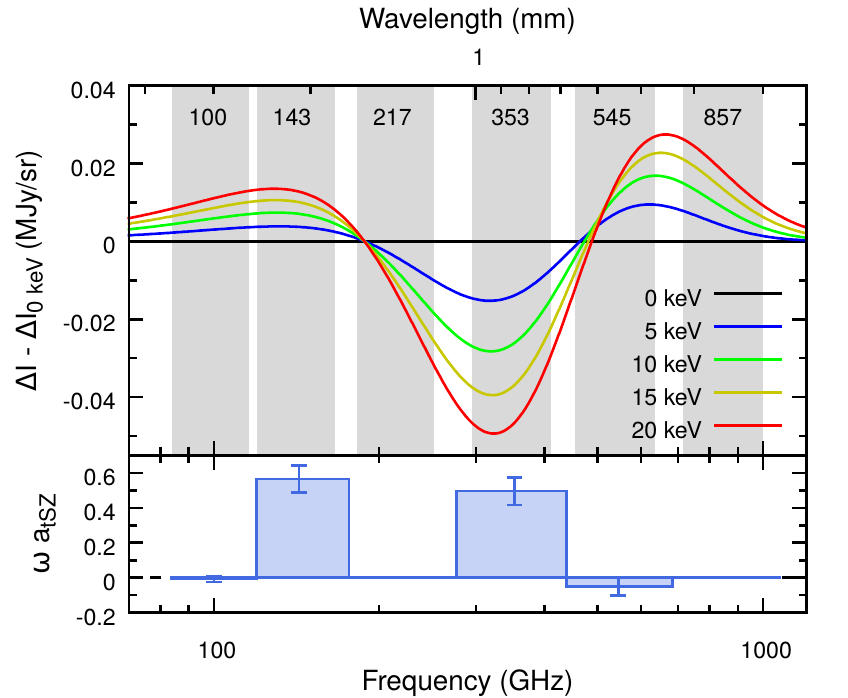}
\caption{Explanation of the high $y$-bias found in ILC $y$-maps. The Top panel shows the difference of the tSZ spectrum computed for different temperatures to the tSZ spectrum at $0 \, \mathrm{keV}$ for $y=10^{-4}$. The Bottom panel shows the product of the ILC-weights $\bm{\omega}$ and the mixing vector of the non-relativistic tSZ spectrum $\bm{a}_\mathrm{tSZ} = \Delta I_\mathrm{tSZ}(0 \, \mathrm{keV})/y$, i.e. the fraction that each channel contributes to the estimated $y$. Summing the product up over all channels will yield $1$. The ILC algorithm produces a high $y$-bias because of the high weights that are assigned to the $143 \, \mathrm{GHz}$ and $353 \, \mathrm{GHz}$ channel at which the difference between the relativistic and non-relativistic tSZ spectra is particularly large.}
\label{fig:ILC_weights}
\end{figure}

The large $y$-bias found in ILC $y$-maps can be understood by computing the contribution of each channel in the linear combination. The contribution is given by the product of the weights $\bm{\omega}$ and the mixing vector $\bm{a}_\mathrm{tSZ}$ for the tSZ, which is shown in Fig.~\ref{fig:ILC_weights} and compared against the difference of the relativistic and non-relativistic tSZ spectrum for different temperatures. It can be seen that the ILC algorithm assigns particularly high weight to the $143 \, \mathrm{GHz}$ and $353 \, \mathrm{GHz}$ channels where the difference between the spectra is particularly high, resulting in the large bias observed in our simulations. We stress that this result is not limited to our simulations and that similar ILC-weights are also found in the official maps made public by the \citet{Planck_YMAPS} that were created using the more sophisticated {\small MILCA} and {\small NILC} algorithms.

\section{Comparison of $T_\mathrm{SZ}$ and $T_\mathrm{X}$}
\label{sec:temperatures}
  \begin{figure}
    \centering
      \includegraphics{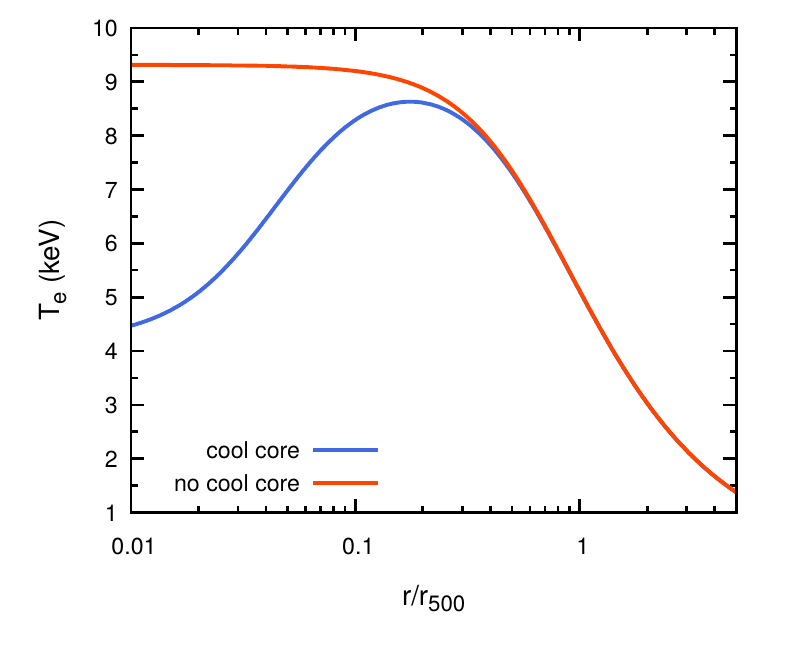}    
    \caption[]{Example cool-core (blue) and non-cool-core (red) radial electron temperature profiles for a cluster with $M_{500} = 6 \times 10^{14} \, \mathrm{M_\odot}$. The cool core profile is taken from \citet{Vikhlinin06} and the non-cool-core one is obtained by slightly modifying it by choosing an infinitesimally small cooling radius.}
    \label{fig:t_profile}
  \end{figure}
\begin{figure}
    \centering
      \includegraphics{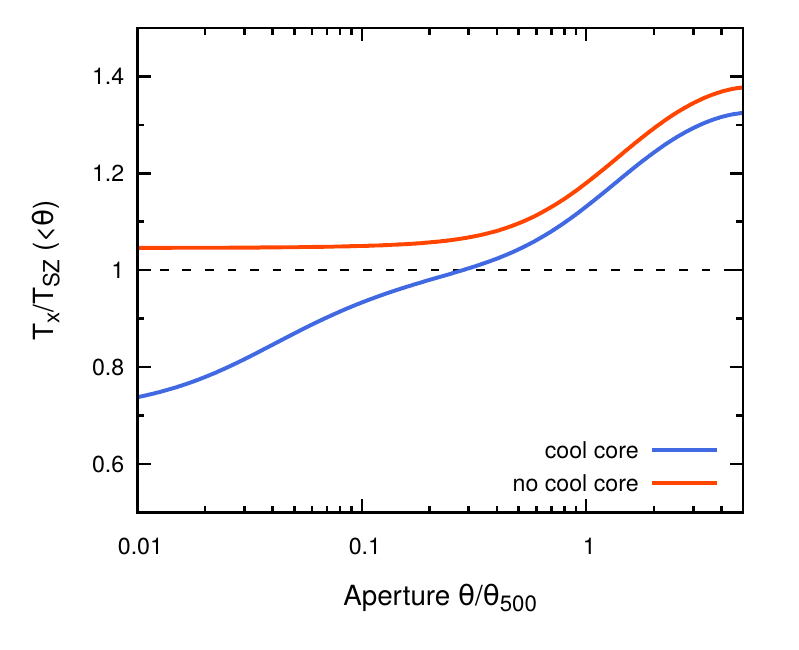}     
    \caption[]{\small Comparison of the expected X-ray spectroscopic and SZ measured ICM temperatures within different apertures. The temperature ratio $T_\mathrm{X}/T_\mathrm{SZ}$ is shown as a function of aperture size $\theta$ assuming a cool-core (blue) and a non-cool-core (blue) $T_\mathrm{e}$-profile. We find that the ratio is always unity for non-cool-core clusters and only smaller then unity for cool-core clusters when $\theta \lesssim 0.3 \theta_{500}$.}
    \label{fig:t_ratio}
  \end{figure}

  We compare the expected aperture-average values for $T_\mathrm{X}$ and $T_\mathrm{SZ}$ using electron pressure and temperature profiles from \citet{Arnaud10} and \citet{Vikhlinin06} that were used to create our mock data sets. The aperture average temperatures are given by 
  \begin{equation}
   T_\mathrm{obs.}(<\theta) = \frac{\int w \, T_\mathrm{e}(r) \, \mathrm{d}V}{\int w \, \mathrm{d}V},
  \end{equation}
  where $w$ is a method-dependent weight and the volume integrals are carried out for a  cylindrical volume that relates directly to the average signal within the aperture $\theta = R/D_\mathrm{A}$. The X-ray spectroscopic temperature $T_\mathrm{X}$ can be obtained using the spectroscopic-like weight \hbox{$w_\mathrm{X}=n_\mathrm{e}^2T_\mathrm{e}^{-\frac{3}{4}}$} (\citealt{Mazzotta04}), while $T_\mathrm{SZ}$ is well approximated by using \hbox{$w_\mathrm{SZ} = n_\mathrm{e}T_\mathrm{e}$} (\citealt{Hansen04}). We compute the ratio $T_\mathrm{X}/T_\mathrm{SZ}$ using analytical temperature and density profiles and without taking into consideration the effect of gas clumping. The temperature model is taken from \citet{Vikhlinin06}, which we consider as the temperature profile of a typical cool-core cluster. We also construct a non-cool-core variant by reducing the size of the cooling radius in the original Vikhlinin et al. model to an arbitrarily small value. These two input temperature profiles are shown in \hbox{Fig.~\ref{fig:t_profile}}. The corresponding density profiles are obtained by dividing our adopted pressure model by these two temperature profiles. Fig.~\ref{fig:t_ratio} then shows the ratio $T_ \mathrm{X}/T_\mathrm{SZ}$ as a function of aperture size for these two types of clusters.
  
  Assuming the aforementioned radial profiles, we find that for clusters without a cool-core the ratio $T_\mathrm{X}/T_\mathrm{SZ}$ will always be larger than unity due to the density-square weighting of the X-ray spectroscopic temperature. The same density-square weighting will result in $T_\mathrm{X}/T_\mathrm{SZ} < 1$ at very small aperture radii ($\theta \lesssim 0.3 \, \theta_{500}$) for cool-core clusters. At the characteristic aperture $\theta_{500}$, we expect $T_\mathrm{X}/T_\mathrm{SZ} \approx 1.1$ for cool-core and $T_\mathrm{X}/T_\mathrm{SZ} \approx 1.2$ for non-cool-core clusters. However, $T_\mathrm{SZ}$ can be larger than $T_\mathrm{X}$ at all radii in case clusters show significant gas clumping as suggested by hydrodynamical simulations \citep{Kay08, Biffi14}. On the other hand, clusters simulated with more recent, improved smooth particle hydrodynamical (SPH) codes show less clumps and smoother gas and temperature distributions \citep{Beck16}.
  %

\bsp	
\label{lastpage}
\end{document}